\documentclass[acmtog,screen,nonacm]{acmart}

\usepackage{booktabs} 

\usepackage{bbm}

\usepackage[ruled]{algorithm2e} 

\SetAlFnt{\small}
\SetAlCapFnt{\small}
\SetAlCapNameFnt{\small}
\SetAlCapHSkip{0pt}

\acmJournal{TOG}

\usepackage[skip=0pt]{caption}
\begin{document}
\title{Neural Marching Cubes}

\author{Zhiqin Chen}
\affiliation{
  \institution{Simon Fraser University}
  \city{Burnaby}
  \country{Canada}}
\email{zhiqinc@sfu.ca}

\author{Hao Zhang }
\affiliation{
  \institution{Simon Fraser University}
  \city{Burnaby}
  \country{Canada}}
\email{haoz@sfu.ca}

\renewcommand\shortauthors{Chen and Zhang}

\begin{abstract}
We introduce {\em Neural Marching Cubes\/}, a {\em data-driven\/} approach for extracting a triangle mesh from a discretized implicit field.
We base our meshing approach on Marching Cubes (MC), due to the simplicity of its input, 
namely a uniform grid of signed distances or occupancies, which frequently arise in surface reconstruction and from neural implicit 
models.
However, classical MC is defined by coarse 
tessellation templates isolated to individual cubes. While more refined tessellations have been proposed by several MC variants, they all make
heuristic assumptions, such as trilinearity, when determining the vertex positions and local mesh topologies in each cube.
In principle, none of these approaches can reconstruct geometric features that reveal coherence or dependencies between nearby cubes (e.g., a {\em sharp edge\/}), as such information is unaccounted for, resulting in poor estimates of the true underlying implicit field.
To tackle these challenges, we 
re-cast MC from a deep learning perspective, by designing tessellation templates more apt at preserving geometric features, and learning the vertex positions and mesh topologies from training meshes, to account for {\em contextual\/} information from nearby cubes.
We develop a compact per-cube parameterization to represent the output triangle mesh, while being compatible with neural processing, so that a simple 3D convolutional network can be employed for the training.
We show that all topological cases in each cube that are applicable to our design can be easily derived using our representation, and the resulting tessellations can also be obtained naturally and efficiently by following a few design guidelines.
In addition, our network learns local features with limited receptive fields, hence it generalizes well to new shapes and new datasets. 
We evaluate our neural MC approach by quantitative and qualitative comparisons to all well-known MC variants.
In particular, we demonstrate the ability of our network to recover sharp features such as edges and corners, a long-standing issue of 
MC and its variants. Our network also reconstructs local mesh topologies more accurately than previous approaches.
Code and data are available at \href{https://github.com/czq142857/NMC}{https://github.com/czq142857/NMC}.

\end{abstract}



\begin{teaserfigure}
  \centering
  \includegraphics[width=1.0\linewidth]{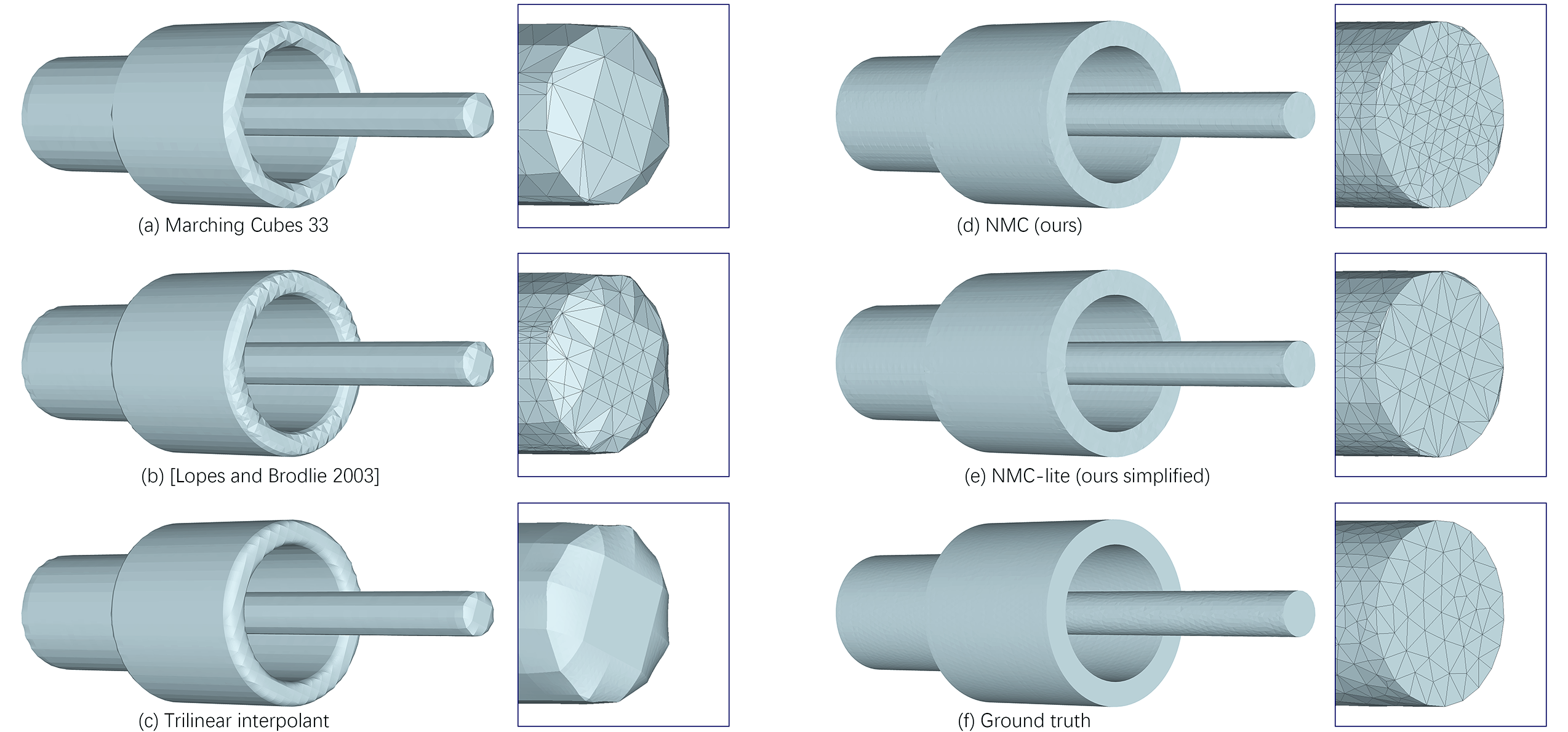}
	\caption{Our Neural Marching Cubes (NMC) is trained to reconstruct the zero-isosurface of an implicit field, while preserving geometric features such as sharp edges and smooth curves. We compare NMC (d), and a simplified version (e), to the best-known MC variants (a-b), as well as a trilinear interpolant (c), none of which could reconstruct the features. The inputs to all methods are the same: a uniform grid of signed distances sampled from the ground truth (f).} 
	\label{fig:teaser}
\end{teaserfigure}

\maketitle

\section{Introduction}
\label{sec:intro}

The Marching Cubes (MC) algorithm \cite{lorensen1987marching} is the most prominent method for isosurface extraction, and has been widely adopted in scientific visualization, shape reconstruction, and by the recent emerging approaches for learning neural implicit representations~\cite{IMNET,OccNet,DeepSDF}. MC takes as input a uniform grid of values representing a discretized implicit field, and extracts a triangle mesh representing the zero-isosurface of the field. The classical MC determines the local mesh topology and tessellation in each cube of the grid by examining the signs at the eight cube corners and referring to a predefined look-up table indexed by the sign configurations. If the isosurface intersects a cube edge, a new mesh vertex is added to that edge with its position computed via linear interpolation. 

With its popularity and wide adoption, MC has seen notable improvements over the years. Marching Cubes 33 \cite{chernyaev1995marching} is one of the first works to assume that the implicit field in each cube follows {\em trilinear interpolation\/} with respect to the cube vertices, and the ensuing meshing algorithm aims for topological correctness under the trilinearity assumption. 
This increases the number of unique cases of mesh tessellations from 15 (in the original MC \cite{lorensen1987marching}) to 33, hence the name. The algorithm itself 
requires many tests to determine which topological case a cube belongs to, and the process can be error-prone. As a result, many patch-ups and improvements have been made to MC 33 \cite{lewiner2003efficient, lopes2003improving, etiene2011topology, custodio2013practical}. Still, the trilinearity assumption, which has persisted, can lead to poor estimates of the true implicit field in general, and especially near {\em sharp features\/} of a shape, as shown in Figure~\ref{fig:teaser}.

Indeed, while MC has been employed for decades, its inability to recover sharp features, arguably its most long-standing issue, has not been
fully resolved. Earlier tessellation templates \cite{lorensen1987marching, DataStructureforSoftObjects,chernyaev1995marching} were
quite coarse and designed for reconstructing soft and smooth objects.
Even with a refined tessellation to possibly represent sharp features in later follow-ups, 
e.g.,~\cite{lopes2003improving}, there is insufficient information in isolated cubes to disambiguate between soft patches and sharp edges,
and this issue is worsened when the inputs are binary occupancies instead of signed distances. 
In general, a shape edge is not a ``point-wise'' feature, but a geometry property that reveals coherence or dependencies over neighboring cubes. Hence, edge prediction should account for that {\em context\/}, yet classical MC algorithms have not used such neighbor information.

\begin{figure}
  \centering
  \includegraphics[width=1.0\linewidth]{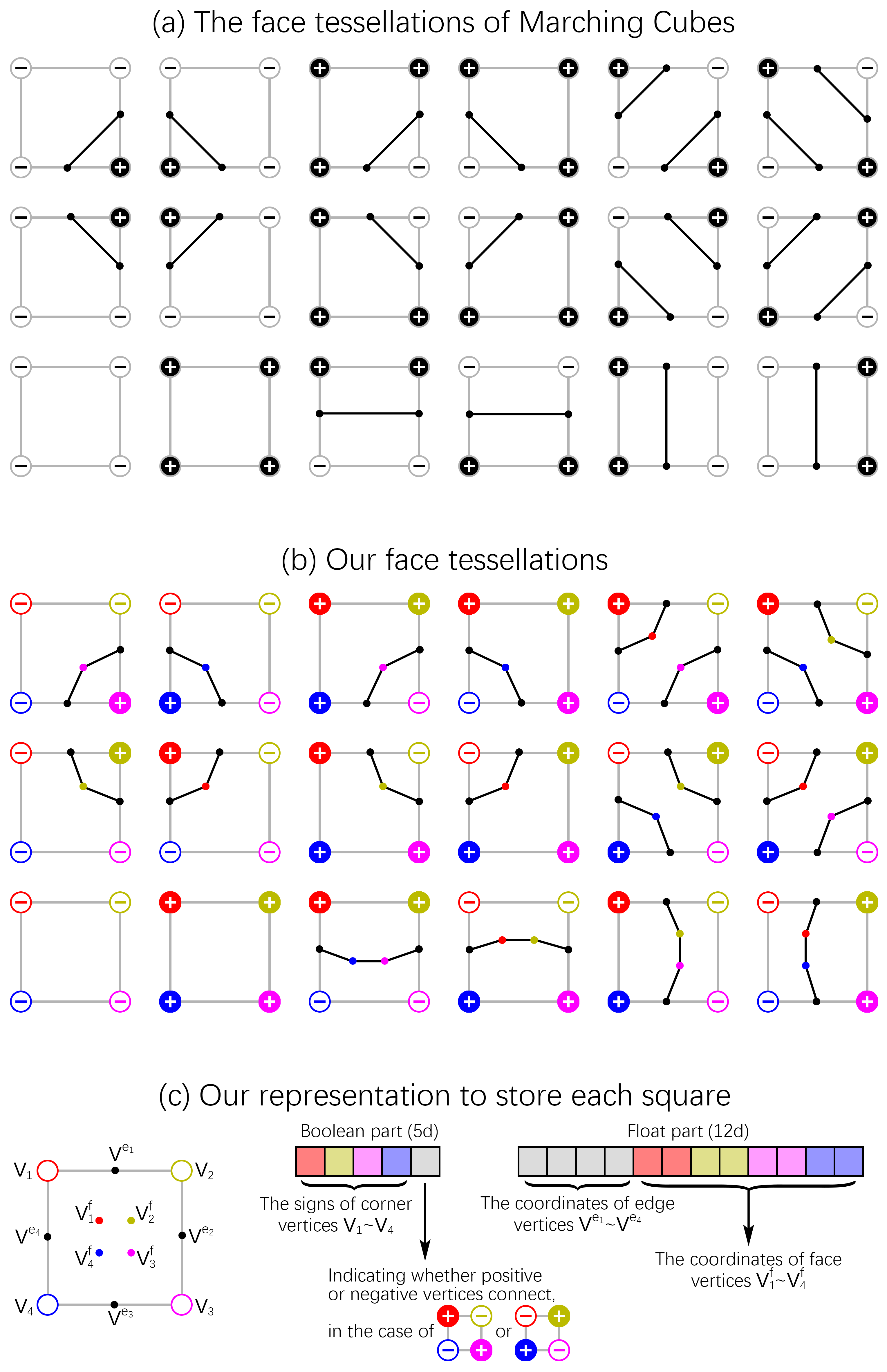}
	\caption{Tessellation design (b) and parameterization (c) for NMC in 2D, in contrast to classical MC (a). Four new (face) vertices are added inside each square (c), each associated to a corner vertex (solid/hollow circle with +/- sign to indicate outside/inside), with matching color. 
Meshing information is encoded by a vector with a ``boolean part" revealing topology and a ``float part'' storing all vertex positions; see Section~\ref{sec:method_2d} for more details.
}
	\label{fig:2D_square}
\end{figure}

In this paper, we introduce {\em Neural Marching Cubes\/} (NMC), a {\em data-driven\/} approach for isosurfacing from a discretized occupancy or signed distance field (SDF). The main premise of our work is that there is sufficient {\em predictability\/} in the vertex positions and local mesh topologies of ``nice'' mesh tessellations under the MC setting, in particular, when they reflect persistent features, such as sharp edges, over neighboring cubes. Hence, a well-designed learning approach would be more effective than handcrafting all the templates and making heuristic decisions such as trilinear interpolation.
To this end, we re-design the tessellation templates so that they are more apt at preserving geometric features including sharp edges and corners, and develop a neural network to learn the vertex positions and mesh topologies from a set of training meshes, so as to account for {\em contextual\/} information from nearby cubes.

To realize NMC, we must address several immediate challenges.
First, we need a per-cube parameterization that is compatible with neural processing, so that the output mesh can be predicted by a network. Such a representation must contain all the information required to perform our modified MC algorithm, including mesh topology and vertex positions in each cell, while minimizing redundancy for efficient network training.
Second, our reconfigured mesh tessellation templates must be {\em refined\/}, {\em complete\/} in the sense of avoiding ambiguities and 
covering all topological varieties, and 
be consistent with our defined representation. 
Last but not least, we must obtain quality training meshes resembling ideal outputs of NMC {\em automatically\/} from a collection of 3D shapes. These ground truth meshes should use the designed tessellation templates in each cube, and be stored in the aforementioned representation.

Figure~\ref{fig:2D_square} illustrates our tessellation design and per-cube representation in the 2D case. In contrast to classical MC, we 
add new vertices to each cube, leading to more refined tessellations that can better represent geometric features.  
By carefully designing the representation (see Section~\ref{sec:method}), all topological cases that are applicable to our design, including all cases in MC 33, as well as all vertex positions, can be compactly encoded in a vector form for neural
processing. Also, by associating the added vertices to their respective corner vertices, the new tessellations can all be obtained 
naturally and efficiently by following a few design guidelines.

With the above representation, our network for NMC is simply a 3D variant of ResNet~\cite{he2016deep} which inputs implicit field values.
The network is trained with ground-truth meshes to set up both the topological and geometric losses, which operate respectively on the binary and float parts of the {\em 3D version\/} (see Figure~\ref{fig:3D_cube_rep}) of the per-cube vector representation shown in Figure~\ref{fig:2D_square}(c).
By limiting its receptive field, our network learns local features, rather 
than from the entire shape, so that it generalizes well to new shapes and new datasets.
Finally, we devise an optimization-based approach (see Section~\ref{sec:method_data}) to obtain the ground truth output meshes from raw 3D shapes based on our representation and tessellation design.

We evaluate NMC by qualitative and quantitative comparisons, in terms of reconstruction quality and feature preservation, to well-known MC variants, on both signed distance and binary voxel inputs.
We show that our method is the first MC-based approach that is able to recover sharp features without requiring additional inputs other than a uniform grid of implicit field values.
In addition, our network can more faithfully reconstruct local mesh topologies 
near thin shape structures and closeby surface sheets.
We also provide a simplified version of NMC, which adopts the same mesh tessellation templates as \cite{lopes2003improving}, to study the fidelity-complexity trade-off. We finally show that our model can be trained to reconstruct clean meshes from noisy inputs by adjusting the training data, thus offering a useful tool for extracting 3D shapes from those shape representations designed for neural networks.

\section{Related work}
\label{sec:related}

Our work is inspired by and closely related to Marching Cubes \cite{lorensen1987marching} and its several variants. For completeness, we also discuss other algorithms for isosurfacing and emphasize the strengths of our NMC approach. Also relevant are recent works from the rapidly advancing field of neural geometry learning.

\subsection{Marching Cubes (MC) and Variants}

The original MC algorithm \cite{lorensen1987marching} was proposed to reconstruct 3D structures from medical scan images, while a concurrent work \cite{DataStructureforSoftObjects} developed similar ideas for reconstructing soft objects. However, these methods did not guarantee surface consistency due to ambiguities of the tessellations in each cube; they may generate holes when tessellations in adjacent cells produce different corner connections on the common face \cite{durst1988re}.
Asymptotic Decider \cite{nielson1991asymptotic} addressed the issue by assuming the implicit field within each face follows bilinear interpolation with respect to the four face vertices, and proposed a solution to produce topologically correct tessellations under the bilinearity assumption.
Several follow-up works \cite{matveyev1994approximation, chernyaev1995marching, nielson2003marching} 
further assumed the implicit field within each cube follows trilinear interpolation with respect to the eight cube vertices.
Specifically, \cite{chernyaev1995marching} was the first work to enumerate all possible topological cases with respect to the trilinear interpolant in the cube, and proposed Marching Cubes 33, which contains 33 unique cases under cube rotational and inversion symmetries (inverting all vertex signs), or 31 cases under rotation, mirroring, and inversion; see Figure~\ref{fig:3D_cube_tess_mc33}(a). In comparison, the original MC algorithm had 15 and 14 cases, respectively.

While correctly enumerating all the topological cases does resolve ambiguities of the tessellations, the many tests required to identify
specific topologies present a computational challenge.
Lewiner et al.~\shortcite{lewiner2003efficient} provided an efficient implementation of MC 33 by utilizing an extended look-up table, but 
left some unresolved issues~\cite{etiene2011topology} which were tackled by follow-up work
\cite{custodio2013practical}. 
Van Gelder and Wilhelms~\shortcite{van1994topological} pointed out that to avoid non-manifold edges, the triangles in the tessellation templates should not lie in the face of a cube, as in MC 33, and this issue was addressed in later improvements 
\cite{cignoni2000reconstruction,lopes2003improving}. 
Specifically, \cite{lopes2003improving} introduced additional vertices on cube faces and interiors, but still followed the trilinearity assumption to place vertices, leading to poor estimates of surface features; see Figure~\ref{fig:teaser}. To extract sharp features from volume data, prior works typically required additional information, such as the positions and normals of the intersection points between cube edges and the shape, for vertex placement \cite{kobbelt2001feature}.
A key point of NMC is that feature recovery is a {\em learnable\/} problem from training meshes. Once trained, our network can accurately predict sharp features and local mesh topologies without any additional input.

\subsection{Other Isosurfacing Algorithms}

Classic isosurfacing algorithms such as dual contouring \cite{ju2002dual} could preserve sharp features, but it
requires the gradients of the intersection points on the edges of the grid cells, which could complicate the input setup. 
Dual contouring could also produce non-manifold edges, which is an issue addressed later by \cite{schaefer2007manifold}. 
In dual marching cubes~\cite{schaefer2004dual}, the implicit function is required as input and queried during reconstruction.
There have also been other extensions to dual contouring~\cite{dey2008delaunay,schreiner2006high}, 
including works~\cite{varadhan2003feature,zhang2004dual} which employ adaptive subdivision for simplification and efficient isosurfacing. Like dual contouring, all these methods require additional inputs such as the gradient information or the function of the underlying implicit field.
By learning from training data, NMC can better preserve geometric features given only a uniform grid of sampled scalar values as input.

Marching Tetrahedra (MT) \cite{doi1991efficient} is another variant of MC: it splits each cube into tetrahedra to produce tessellations therein. The tessellation cases for MT are simpler than thoses of MC, but they were also not designed to recover sharp features. Finally, in a recent work called Analytic Marching \cite{lei2020analytic}, rather than taking a signed distance field to mesh, the input is 
implemented as a trained multi-layer perceptron (MLP) with rectified linear units (ReLU). The meshing is then performed by
a marching over ``analytic cells'', which correspond to linear regions resulting from a partitioning by the MLP. Overall, none of the above isosurfacing algorithms learn mesh tessellations from training data.

\subsection{Neural Geometry Learning}

With rapid advances in geometric deep learning, different neural representations have been proposed for 3D shapes. Voxels \cite{3DR2N2,3DGAN,HSP}, point clouds \cite{PCGAN,pointSetGen}, 
and implicit models \cite{IMNET,OccNet,DeepSDF} are among the most popular. But they all require a post processing step to extract a mesh. Deformable meshes/patches \cite{atlasnet,pixel2mesh,3DN} can directly output well-tessellated meshes, but they rely heavily on the input mesh templates and are unable to alter their topologies. There are only a handful of works \cite{BSPNET,SDM,cvxnet,scan2mesh} 
that could output polygonal meshes directly. In contrast, our method takes a discretized implicit field as input and directly outputs a triangle mesh, and therefore could act as the post processing step for many of the above representations.

NMC follows a recent trend in applying machine learning to classical low-level geometry processing tasks including mesh denoising~\cite{wang2016}, shape transform~\cite{LOGAN}, point cloud upsampling~\cite{PUNET}, skeletonization~\cite{lin2021skel}, and subdivision~\cite{NeuralSubdivision}, among many others.
In particular, in neural subdivision, Liu et al.~\shortcite{NeuralSubdivision} proposed a graph neural network to perform geometry-aware subdivision on triangle meshes. It recursively subdivides an input mesh by applying classic loop subdivision, while the positions of the newly added vertices are predicted by a neural network conditioned on local geometry.

In terms of combining machine learning and MC, two notable works, Deep Marching Cubes (DMC) \cite{DeepMarchingCubes} and MeshSDF \cite{MeshSDF}, both aim to make MC {\em differentiable\/}.
Specifically, DMC learns a differentiable approximation of MC by predicting probabilistic occupancies and vertex displacements, while MeshSDF
adopts a continuous model of how signed distance function perturbations locally impact surface geometry to obtain a differentiable surface parameterization.
Another work, DefTet\cite{gao2020deftet}, shares a similar spirit as DMC, as it reconstructs tetrahedral meshes by predicting occupancies in an initial tetrahedral grid, and deforming the vertices to approximate the output shape.

Our work differs from DMC and MeshSDF in several major ways. First, our goals are different. The goal of DMC and MeshSDF is to directly obtain an explicit mesh representation from discrete raw inputs, e.g., point clouds, voxels, or images, in an end-to-end trainable manner, while NMC builds a framework to make MC learnable from training meshes. The input to NMC is a discrete implicit field of distances or occupancies obtained by any model, neurally or not. Second, the focus of DMC and MeshSDF is the end-to-end differentiability, while our focus is on designing and training a refined neural MC model to better reconstruct geometry and topology, in particular sharp features, unlike any other previous MC variant or differential extension. Case in point, DMC only adopted 8 out of the 15 mesh tessellation templates from the original MC~\cite{lorensen1987marching} that have singly connected topologies, falling far short of~\cite{lopes2003improving} and NMC in terms of topological granularity. Last but not the least, DMC and MeshSDF rely on global features to predict the output shapes, not aiming to generalize to other shape categories not present in the training set. In contrast, our network employs a limited receptive field for each cube, leading to a more robust and general isosurfacing algorithm.

\section{Neural Marching Cubes}
\label{sec:method}

We detail our representation for performing Neural Marching Cubes (NMC). We first introduce in a 2D example how the local topologies and tessellations in a square can be represented with a fixed-length code of binary values and float numbers; see Figure~\ref{fig:2D_square}. Then we expand the representation into the 3D cube for NMC, as shown in Figure~\ref{fig:3D_cube_rep}. 
We show how to design the mesh tessellations with respect to our representation and how our training data could be generated from raw meshes. Finally, we describe the network architecture and objective functions we designed to train NMC.

\subsection{2D NMC: representation in a 2D square}
\label{sec:method_2d}

We follow the common assumption in MC that if the two vertices of an edge in a cube (or a square) have different signs, there will be one and only one intersection point between the edge and the underlying zero-isosurface. As a result, all the situations in a square can be enumerated as in Figure~\ref{fig:2D_square}(a). However, the tessellation templates in the classic MC algorithms are unable to represent geometric features such as sharp edges by design, hence they must be re-designed. Simply adding one vertex on each generated edge (black line) of the original templates could already solve the issue. But since we want a neural network to fully predict the meshing in each square, including the added vertices, we need to design a representation with a fixed format to store all the necessary information, so that the representation could be directly parsed into an output mesh, while being compatible with neural processing and training.

First, we add four face vertices ($\textrm{v}_\textrm{1}^\textrm{f} \sim \textrm{v}_\textrm{4}^\textrm{f}$) on the face, each associated with one corner vertex ($\textrm{v}_\textrm{1} \sim \textrm{v}_\textrm{4}$), as shown in Figure~\ref{fig:2D_square}(c) left. With the added vertices, new tessellations that can better preserve geometric features can be easily derived; see Figure~\ref{fig:2D_square}(b).

Next, we design a fixed-length vector to fully encode the output mesh (edges) in each square; see Figure~\ref{fig:2D_square}(c). The vector is split into a {\em boolean\/} or {\em binary part\/} to describe the topological cases, and a {\em float part\/} to store floating point numbers as vertex positions.

When the signs of the four corner vertices are given, there is only one ambiguity case, when both ends of a diagonal line are with the same signs but the ends of an edge are with different signs; see top-right corner of Figure~\ref{fig:2D_square}(b). This ambiguity can be resolved by adding a face sign which is positive if the connected vertices are positive, and vice versa. Hence, the boolean part has $5$ values storing the signs: one face sign and four signs for the corner vertices. On the other hand, we need to store all vertex positions in the float part, whether the vertices are being used or not. Since each edge vertex only has one degree of freedom, four edge vertices take $4$ floats to store. Adding the $8$ numbers for the 2D coordinates of the four face vertices, in total we have $12$ numbers in the float part.

However, note that the representation for each square is not the same as the representation we use to predict the entire 2D shape, because of the redundancy: an edge vertex is shared by two adjacent squares, and a corner vertex is shared by four. Therefore, when representing the squares of an entire shape, we only need to store the sign of one corner vertex ($\textrm{v}_\textrm{1}$) and the face sign in the boolean part, and two edge vertices ($\textrm{v}^{\textrm{e}_\textrm{1}}$,$\textrm{v}^{\textrm{e}_\textrm{4}}$) and all four face vertices in the float part, leading to 2d and 10d vectors, respectively. Afterwards, a 2D Convolutional Neural network (CNN) could be applied to take as input an $M \times N$ array of implicit field values, and output an $M \times N \times 12$ array that is parsed into a 2D mesh.

\subsection{3D NMC: representation in a 3D cube}
\label{sec:method_3d}

\begin{figure*}[!th]
  \centering
  \includegraphics[width=0.99\linewidth]{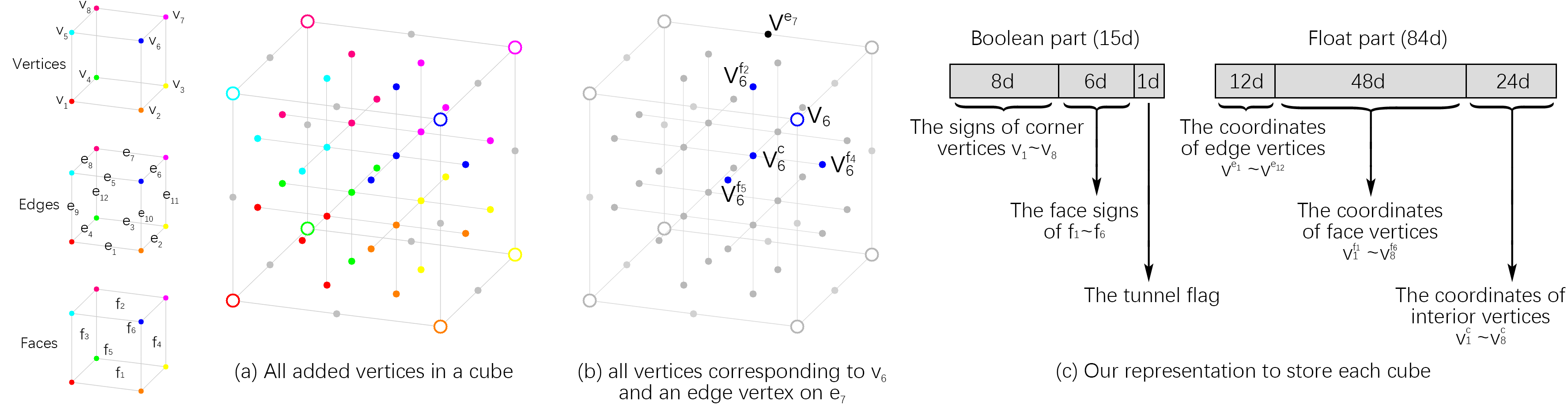}
	\caption{Per-cube parameterization for our NMC in 3D, with 12 edge vertices ($\textrm{v}^\textrm{$\textrm{e}_\textrm{i}$}$), $6\times 4=24$ new face vertices ($\textrm{v}_\textrm{j}^\textrm{$\textrm{f}_\textrm{k}$}$), along with 8 new interior vertices ($\textrm{v}_\textrm{j}^\textrm{c}$), as shown in (a). Vertices with the same color correspond to the same cube vertex, as shown in (b), where the grey lines in (a-b) are for ease of visualization only. The vector representation for local mesh topology (the boolean part) and vertex positions (the float part) is given in (c), where the number of floats needed to represent a vertex depends on the degrees of freedom, e.g., one for an edge vertex, two for a face vertex, and three for an interior vertex.}
	\label{fig:3D_cube_rep}
\end{figure*}

As shown in Figure~\ref{fig:3D_cube_rep}, we design the NMC representation for a 3D cube in a similar manner as its 2D counterpart shown in Figure~\ref{fig:2D_square}. In addition to the edge vertices ($\textrm{v}^\textrm{$e_1$} \sim \textrm{v}^\textrm{$e_{12}$}$), we keep the four added face vertices for each of the six faces of the cube. We also add eight interior vertices ($\textrm{v}_\textrm{1}^\textrm{c} \sim \textrm{v}_\textrm{8}^\textrm{c}$) in the cube, each affiliated with one corner vertex ($\textrm{v}_\textrm{1} \sim \textrm{v}_\textrm{8}$) of the cube. Details on how to properly tessellate the cube with these new vertices will be discussed in Section~\ref{sec:method_3dt}. In this section, we focus on how to represent the topological cases and the positions of the added vertices, using a boolean and a float part respectively, as shown in  Figure~\ref{fig:3D_cube_rep}(c).

For the boolean part, we require at least eight signs at the corner vertices of the cube and six face signs to describe or index the local mesh topology in each cube. However, these are not sufficient to resolve all ambiguities. As already observed in
MC 33~\cite{chernyaev1995marching}, 
when there are two connected components with the same sign on the surface of a cube, the two could be connected with a tunnel inside the cube. The real situations could be far more complicated than that. There could be more than two connected components on the surface of the cube, and there could be more than one tunnel to connect the two components. To simplify the situation, we draw inspiration from the topological cases in MC 33, which are shown in Figure~\ref{fig:3D_cube_tess_mc33}(a). Note that all the 33 cases have either zero or one tunnel, if there are exactly two connected components with the same sign. Therefore, we assume that in the case of two connected components, there could be one tunnel connecting the components, or none at all. In other cases with just one or more than two connected components, we assume zero tunnel. With such a simplifying assumption, we only need to add one binary value to indicate whether a tunnel exists, leading to a total of $15$ binary values in the boolean part.

For the float part, we need to store $44$ added vertices: $12$ edge vertices, $6 \times 4 = 24$ face vertices, and $8$ interior vertices, as shown in Figure~\ref{fig:3D_cube_rep}(a-b). Since each edge vertex only has one degree of freedom, $12$ edge vertices would only need $12$ floats to store. Each face vertex has two degrees of freedom, therefore $24$ face vertices take $48$ floats to store. Plus the $24$ floats for the 3D coordinates of the $8$ interior vertices, in total we have $84$ numbers in the float part.

However, similar to the 2D cases, the representation for each cube is not the same as the representation we use to predict the entire 3D shape, because of the redundancy: a corner vertex is shared by eight cubes, an edge vertex is shared by four, and a face vertex is shared by two. Therefore, when representing the cubes of an entire shape, we only need to store $5$ values in the boolean part (the sign of $\textrm{v}_\textrm{1}$, $\textrm{f}_\textrm{1}$, $\textrm{f}_\textrm{3}$, $\textrm{f}_\textrm{5}$; and the tunnel flag), and $51$ values in the float part ($3$ edge vertices on edges $\textrm{e}_\textrm{1}$, $\textrm{e}_\textrm{4}$, $\textrm{e}_\textrm{9}$; $12$ face vertices on faces $\textrm{f}_\textrm{1}$, $\textrm{f}_\textrm{3}$, $\textrm{f}_\textrm{5}$; and all $8$ interior vertices). Afterwards, a 3D CNN can be applied to take as input an $M \times N \times P$ array of implicit field values, and output an $M \times N \times P \times 56$ array that could be parsed into a 3D mesh.

\subsection{3D NMC: tessellating a 3D cube}
\label{sec:method_3dt}

\begin{figure*}
  \centering
  \includegraphics[width=1.0\linewidth]{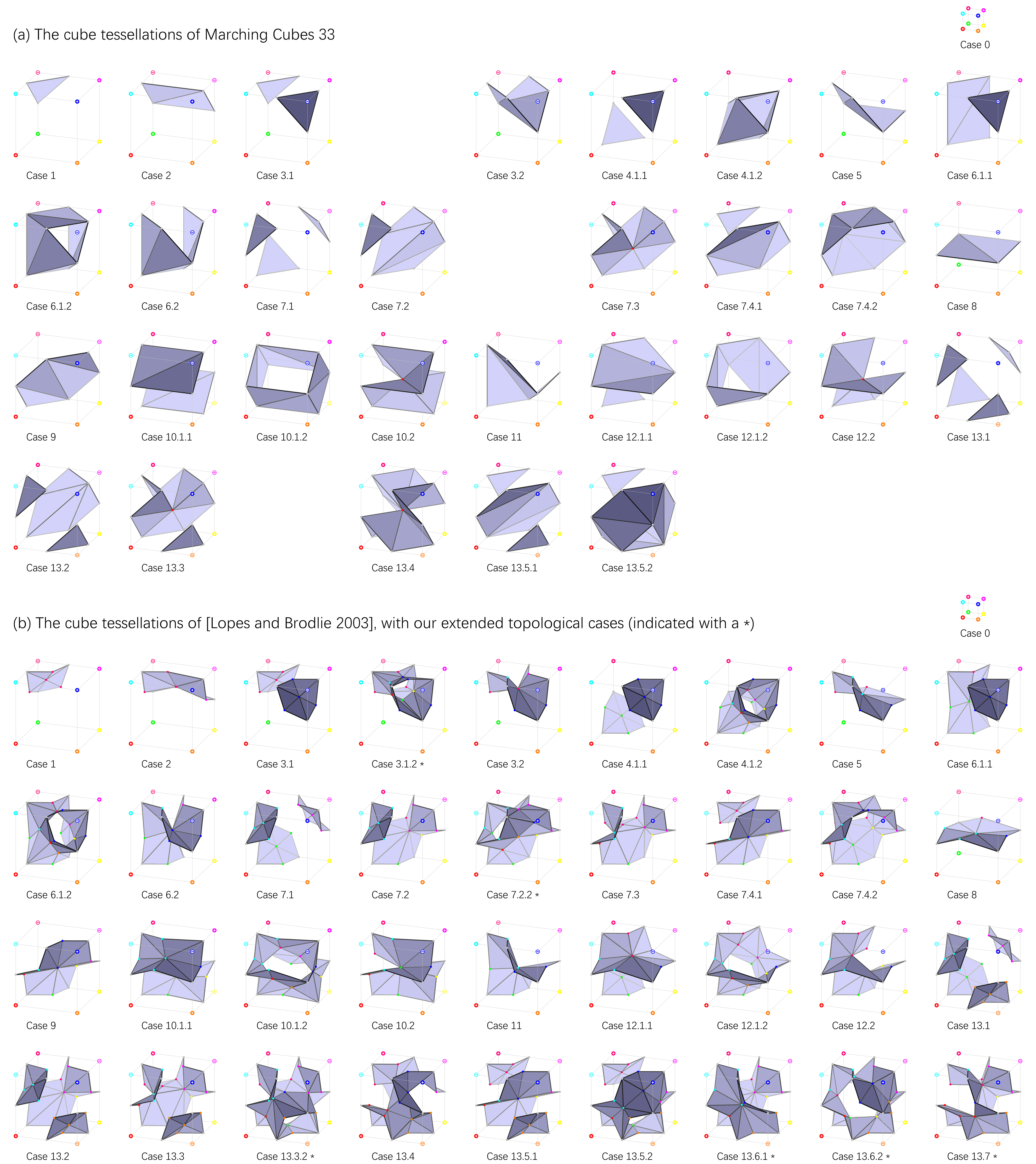}
	\caption{The 3D cube tessellations of Marching Cubes 33~\cite{chernyaev1995marching} and~\cite{lopes2003improving}. Note that they both present 31 cases, since Case 12.3 is equivalent to Case 12.2 and Case 14 is equivalent to Case 11, with respect to rotational and mirroring symmetries. In (b), we also add our extended topological cases to~\cite{lopes2003improving}, indicated with a *, to form a simplified version of our NMC tessellations, denoted as NMC-lite.}
	\label{fig:3D_cube_tess_mc33}
\end{figure*}
\begin{figure*}
  \centering
  \includegraphics[width=1.0\linewidth]{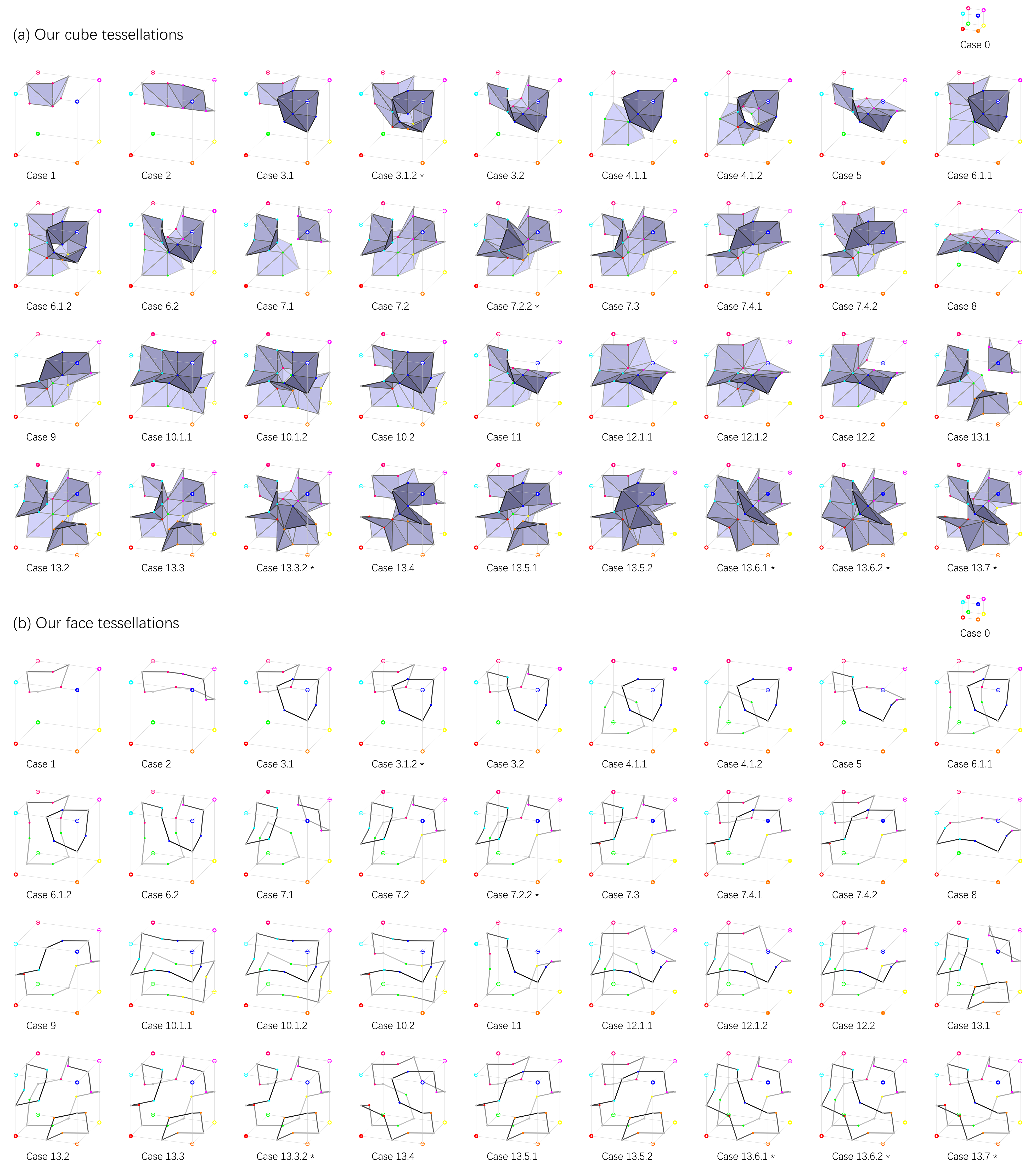}
	\caption{Our cube tessellations and face tessellations for all the 37 unique topological cases considered by NMC, where vertices with the same color correspond. Note Case 0 in the top-right corner which indexes the case where all signs on the cube vertices are the same.}
	\label{fig:3D_cube_tess_ours}
\end{figure*}
\begin{figure}
  \centering
  \includegraphics[width=1.0\linewidth]{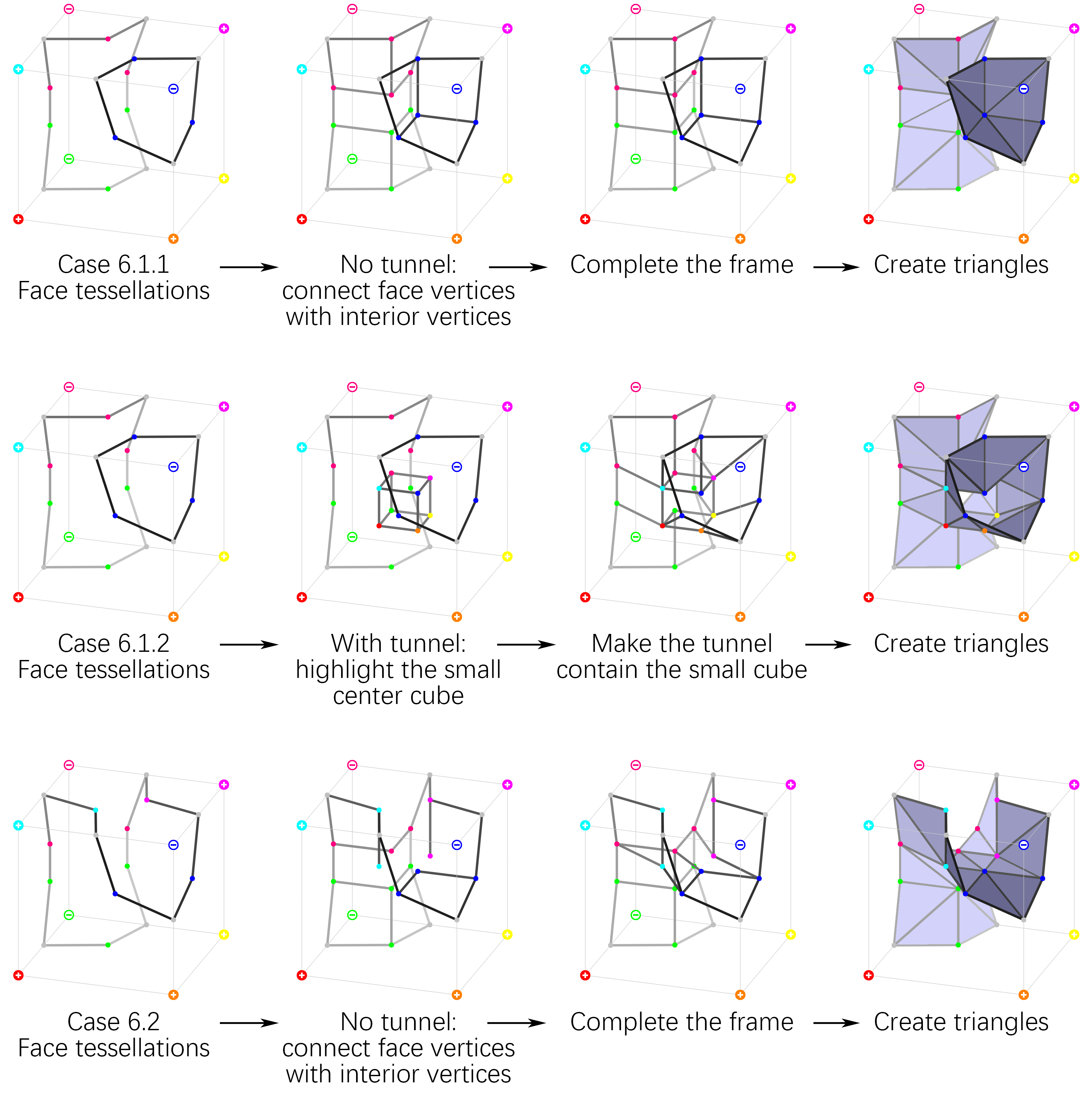}
	\caption{Example tessellation steps for our NMC designs. The face tessellations in the first column follow Figure~\ref{fig:2D_square}, therefore they are considered as ``given'', and we only need to add new structures inside the cube.}
	\label{fig:3D_cube_tess_demo}
\end{figure}

In this section, we elaborate how we tessellate the cube with respect to each topological case. To facilitate the tessellation design, we developed a graphical user interface for interactive modelling, and employed the interface to design and render all the cases as shown in Figures~\ref{fig:3D_cube_tess_mc33} and~\ref{fig:3D_cube_tess_ours}.

Generally, the tessellation design needs to comply with several basic principles. First, the resulting mesh should contain only triangles. Second, the mesh should completely separate vertices with different signs, i.e., any path inside the cube that connects two vertices of different signs must intersect with the mesh. Third, there should be no non-manifold edges or vertices. Specific to our method, there is a fourth principle to follow: we are only allowed to use the vertices present in our cube representation as described in Section~\ref{sec:method_3d}.

However, the very first step we need to take before designing the tessellations, is to enumerate how many unique topological cases there are. In our cube representation, we have $15$ binary values to describe the cases, therefore we have a total of $2^{15}=32,768$ cases. Yet, if we consider the presence of rotational symmetries, mirroring symmetries, inversion symmetries (inverting all vertex and face signs in a cube), and remove all invalid cases with respect to the tunnel flag, we have a total of {\em {\bf 37} unique cases.}

We can directly use the face tessellations in Figure~\ref{fig:2D_square} to tessellate the six faces of a cube, as shown in Figure~\ref{fig:3D_cube_tess_ours}(b). Afterwards, we follow several guidelines to create a mesh inside the cube: a) If there is a tunnel, then the tunnel must contain the small center cube made by the eight interior vertices; b) if there is no tunnel, then connect all available face vertices to their corresponding interior vertices; c) avoid long triangles. See Figure~\ref{fig:3D_cube_tess_demo} for several examples.

The completed tessellations of our method can be found in Figure~\ref{fig:3D_cube_tess_ours}(a). The tessellation design allows much freedom and does not necessarily have to follow our guidelines. For example, we could simply take the tessellations in \cite{lopes2003improving} into our framework, as show in Figure~\ref{fig:3D_cube_tess_mc33}(b). Since this tessellation design employs fewer vertices and triangles, we coin our Neural Marching Cubes using this specific tessellation design as {\em NMC-lite}. Note that in both cases, the training data will be prepared and the network will be trained with respect to their own tessellation designs.

\subsection{Data preparation}
\label{sec:method_data}

\begin{figure}
  \centering
  \includegraphics[width=1.0\linewidth]{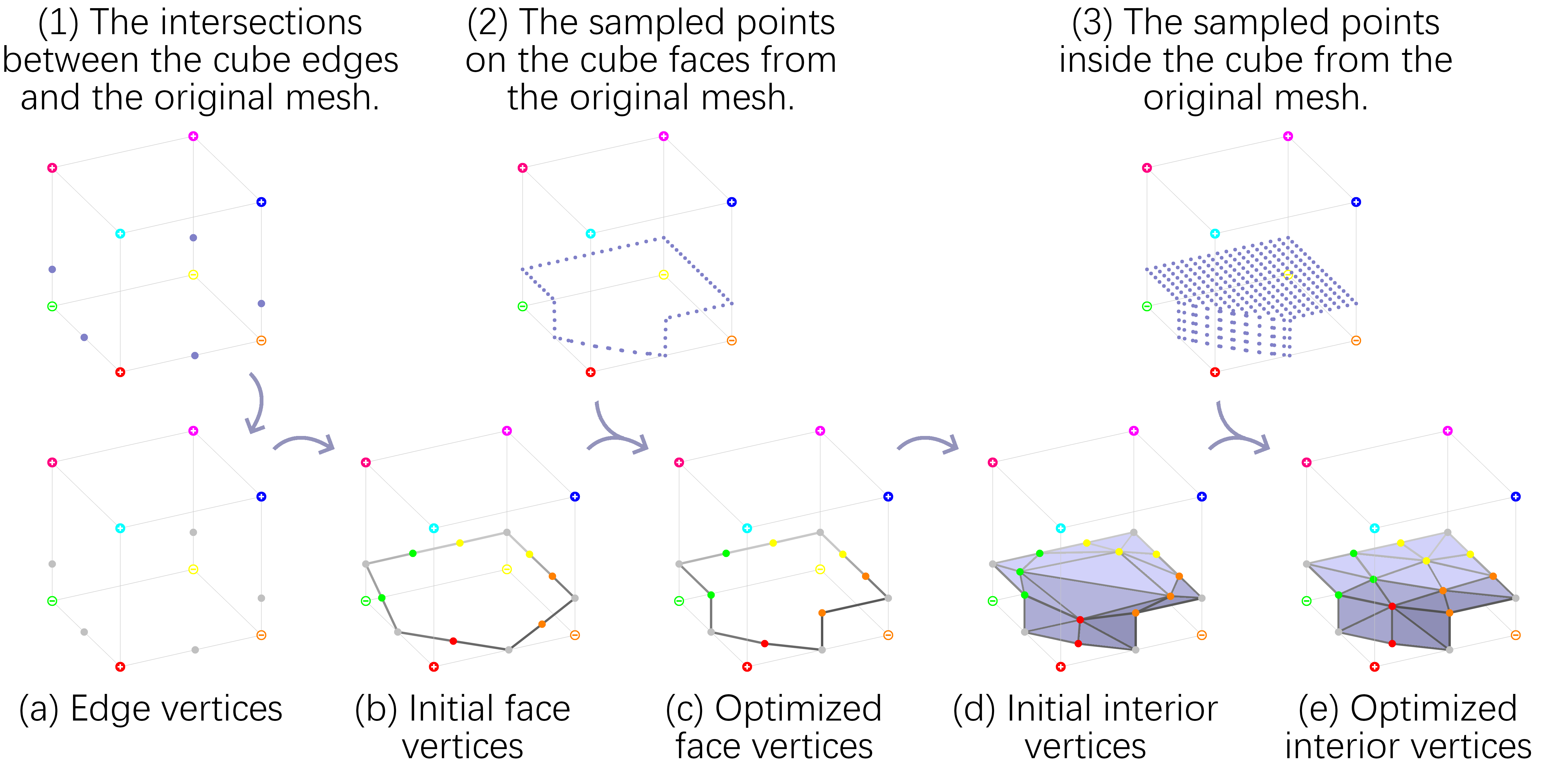}
	\caption{Our preprocessing step to prepare the training mesh data for NMC. After determining the topological case for the cube, we optimize the vertex positions to approximate the original mesh. The initial face vertices are mid-points or trisection points, while the initial interior vertices in the cube are averages of connected edge vertices and face vertices.}
	\label{fig:data_overview}
\end{figure}

We now introduce data preparation for NMC, i.e., the preprocessing step to convert a raw mesh into a form compatible with our cube representation in Figure~\ref{fig:3D_cube_rep} for neural processing; it is an $M \times N \times P \times 5$ array for the boolean part and an $M \times N \times P \times 51$ array for the float part, when the input grid is $M \times N \times P$. We divide a raw 3D mesh into small cubes to process each separately, as in MC. For each cube, we first determine its topological case. Then we put the corresponding tessellation template inside that cube, and optimize the vertices of the tessellation template to minimize the Chamfer Distance with respect to the original mesh. An overview is given in Figure~\ref{fig:data_overview}.

To determine the topological case in a cube, we compute a $9 \times 9 \times 9$ grid of signed distances inside the cube, so that each face contains $9 \times 9$ signed distances. We then check the connectivities between the vertices through the SDF grid, where adjacent grid points with the same sign are considered connected, to determine the case for each of the six faces. After the face cases are determined, we only need to test whether there is a tunnel to determine the cube case, which can be done by checking the number of connected components inside the cube. Note that in several situations, the cube cannot be represented with our templates, e.g., when there are two or more intersections on a cube edge, or when there are complex structures inside the cube that are unaccounted for. 

We perform tests to validate whether an edge/face/cube can be represented using our templates by checking the number of connected components, which are compared against the templates in Figure~\ref{fig:2D_square} (for faces in the 2D case) or Figure~\ref{fig:3D_cube_tess_ours} (for cubes in 3D). The edges/faces/cubes that do not have matching numbers are deemed to be invalid.
For example, if the end vertices (grid points) of an edge are with different signs, then the $9$ grid points on the edge should contain exactly two connected components, one positive and one negative. In a 3D cube, say Case 6.1.1, there are three connected components, one positive and two negatives, while in Case 6.1.2, there are two connected components because of the tunnel.

The removal of invalid edges/faces/cubes from the training meshes is critical to NMC and this is accomplished by a {\em masking\/} process.
Specifically, we generate masks during data preparation to indicate valid values in our representation with {\bf 1's} and invalid or irrelevant values with {\bf 0's}, where invalidity implies that the edge/face/cube cannot be represented by our designed tessellation templates, and a value stored in our representation is irrelevant if it does not affect the output mesh (e.g., the face sign in an unambiguous face, or the tunnel flag in a cube that cannot form a tunnel).
For shape $s$, we denote the input array as $I_{s} \in \mathbb{R}^{M \times N \times P}$, the array of the boolean part as $B_{s} \in \{0,1\}^{M \times N \times P \times 5}$, and the array of the float part as $F_{s} \in [0,1]^{M \times N \times P \times 51}$. Therefore, the mask of $B_{s}$ is $M_{B_{s}} \in \{0,1\}^{M \times N \times P \times 5}$ and the mask of $F_{s}$ is $M_{F_{s}} \in \{0,1\}^{M \times N \times P \times 51}$.

After the topological case is settled, we put the corresponding tessellation template inside the cube and optimize its vertices to approximate the original mesh. However, since adjacent cubes share edges and faces, we first determine the positions of all edge vertices, then all face vertices, and finally all interior vertices, to avoid repeated computations. Note that only the edge vertices do not require optimization since we can find them by checking the intersection points between the cube edges and the original mesh. Take the interior vertices for example -- while the face vertices can be optimized in a similar way, we densely sample points on the mesh inside the cube to obtain a point cloud $P$. Denote the vertices, edges, and triangles in the tessellation template as $V$, $E$, and $T$, respectively. Denote the point-triangle Euclidean distance as $D(p,t)$, and the point-point Euclidean distance as $d(v_1,v_2)$, we have the objective function
\begin{equation}
L_{total} = L_{P \to T} + L_{T \to P} + \gamma L_{reg}, \text{with}
\end{equation}
\begin{equation}
L_{P \to T} = \frac{1}{|P|} \sum_{p \in P} \min_{t \in T} D(p,t),
\end{equation}
\begin{equation}
L_{T \to P} = \frac{1}{|T|} \sum_{t \in T} \min_{p \in P} D(p,t),
\end{equation}
\begin{equation}
L_{reg} = \frac{1}{|V|} \sum_{v_1 \in V} \min_{\{v_1,v_2\} \in E} d(v_1,v_2),
\end{equation}
where $L_{P \to T}$ is the point-to-triangle distance, $L_{T \to P}$ the triangle-to-point distance, and $L_{reg}$ a regularization term to keep the surface as ``tight'' as possible by minimizing edge lengths, and $\gamma$ is set to $0.1$.

Note that it is possible to use the above objective function to train the network directly, instead of using a mean squared error loss as we will describe in the next section. However, to ensure the quality of the generated mesh, we usually sample a very dense point cloud to peform the optimization. The computational time and memory cost make it intractable to train the network directly with the optimization objectives.

\subsection{NMC network and objective functions}
\label{sec:method_nn}

The input to our network is an array of implicit field values $I_{s}$, and the ground truth outputs contain a boolean array $B_{s}$ and a float array $F_{s}$. The masks $M_{B_{s}}$ and $M_{F_{s}}$ indicate which values in $B_{s}$ and $F_{s}$ are valid. Since they are all 3D arrays (with feature channels), we could directly apply 3D convolutional neural networks for the task. Specifically, we use a 3D variant of ResNet~\cite{he2016deep} as our backbone network, with receptive fields of size $7^3$. 

For the objective functions, we use a binary cross entropy (BCE) loss for the boolean part and a mean squared error (MSE) loss for the float part. Denote the outputs of our network as $D_{s} = f_{B}(I_{s})$ and $H_{s} = f_{F}(I_{s})$ for the boolean part and float part, respectively, and denote the entire shape dataset as $\mathcal{S}$. Let all multiplications in the following equations be element-wise products, then we have
\begin{equation}
L_{bool} = \mathbb{E}_{s \sim \mathcal{S}} || - M_{B_{s}}( B_{s} \log(D_{s}) + (1-B_{s}) \log(1-D_{s}))||_1,
\end{equation}
\begin{equation}
L_{float} = \mathbb{E}_{s \sim \mathcal{S}} || M_{F_{s}}( F_{s}-H_{s})||_2^2.
\end{equation}
We could directly sum $L_{bool}$ and $L_{float}$ with a weighting parameter to obtain the final objective function. However, our experiments showed that it is tedious to find an appropriate weight for the two terms. Therefore, we choose to use two distinct networks to predict $D_{s}$ and $H_{s}$ separately, so that one network is trained with $L_{bool}$ and another with $L_{float}$ without any interference.

\begin{figure}
  \centering
  \includegraphics[width=0.9\linewidth]{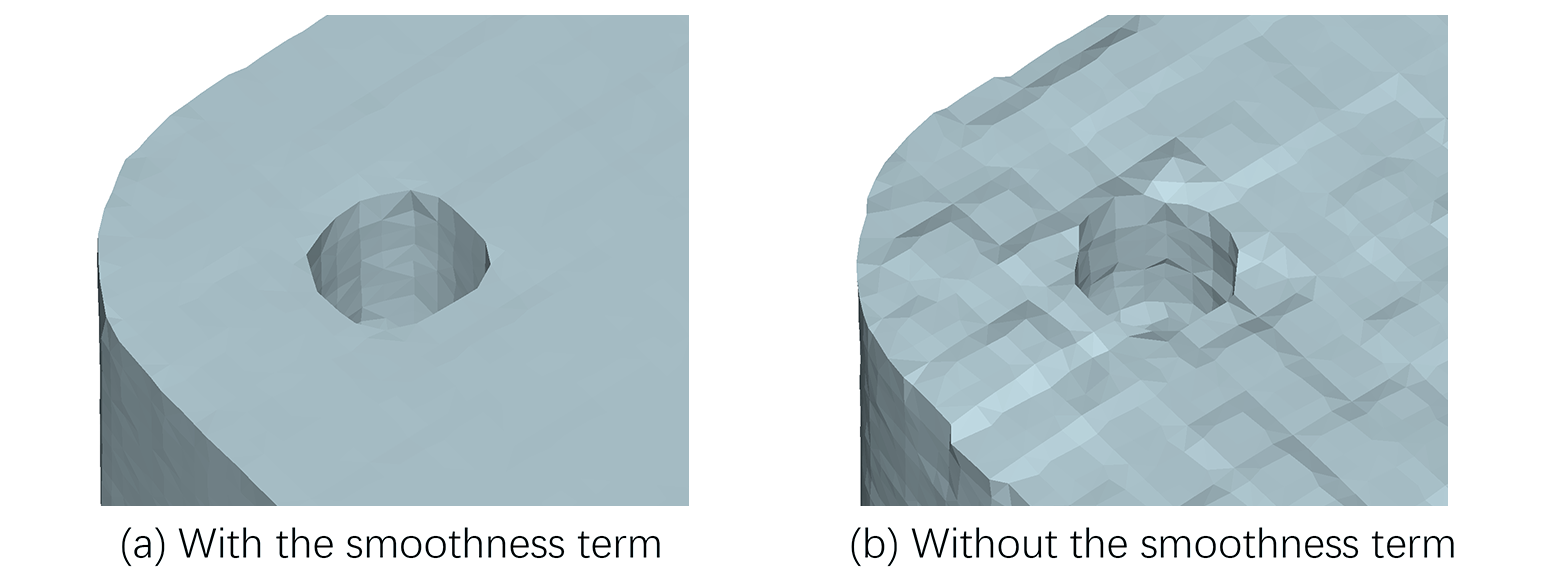}
	\caption{Output meshes when our network is trained with vs.~without the smoothness term when the inputs are binary voxel/occupancy grids.}
	\label{fig:smoothness_term}
\end{figure}

However, the above settings are not sufficient for binary voxel inputs, due to considerable ambiguities of the possible shapes represented by the input voxels. Therefore, we use the aforementioned settings for SDF grid inputs, and make a few adjustments when the inputs are binary voxels. Specifically, we enlarge the receptive fields of our backbone network from $7^3$ to $15^3$ to reduce ambiguity, and add a smoothness term to the loss function on the float part,
$$
L_{float}^{*} = L_{float} + \gamma L_{smooth}, \text{with}
$$
\begin{equation}\label{eq:smoothness}
L_{smooth} = \mathbb{E}_{s \sim \mathcal{S}} \sum_{(u,v) \in E_s} || \mathbbm{1}(|F^u_s-F^v_s|<\sigma) \cdot (H^u_s-H^v_s) ||_2^2,
\end{equation}
where $E_s$ denotes the set of all edges $(u,v)$ in the ground truth (GT) output mesh for shape $s$, $F^u_s \in \mathbb{R}^{3}$ is the coordinates of vertex $u$ in the GT mesh, and $H^u_s \in \mathbb{R}^{3}$ is the coordinates of $u$ in the {\em predicted\/} mesh. Note that the mesh tessellations of the GT mesh and the predicted mesh are the same since the tessellations are determined solely by the boolean part, and we use the GT boolean part when training the float part. In addition, $\mathbbm{1}(\cdot)$ in the equation converts true/false into $1$/$0$, respectively, and the parameters $\sigma=0.0002$ and $\gamma=10$ are fixed throughout our experiments.

Overall, the smoothness term encourages the output surfaces to align with the {\em coordinate axes\/}, with the underlying assumption that the GT surfaces generally share the same characteristic. We show the impact of $L_{smooth}$ in Figure~\ref{fig:smoothness_term} and explain this choice in the Appendix by experimenting with different smoothness terms.

The network architectures and more training details can also be found in the supplementary material.

\section{Results and evaluation}
\label{sec:results}

In this section, we show results and evaluate NMC both qualitatively and quantitatively, on both SDF and binary voxel inputs. We compare NMC to well-known MC variants, and demonstrate its generalizability and the ability to handle noisy input.

\paragraph{Datasets.}
For our experiments, we mainly work with the first chunk of the ABC dataset~\cite{koch2019abc}, which consists of triangle meshes of CAD shapes. Such CAD shapes are characterized by their rich geometric features (e.g., sharp edges, corners, smooth curves, etc.) and topological varieties. We split the set into 80\% training (4,280 shapes) and 20\% testing (1,071 shapes). During data preparation, we obtain triangle meshes over $32^3$ and $64^3$ grids to train our network. Evaluation is conducted on the testing set.
Later, to assess the generalizability of NMC, we also test (not train) the same network on the Thingi10K dataset \cite{zhou2016thingi10k}, which contains a variety of 3D-printing models.

\paragraph{Evaluation metrics.}
To perform quantitative evaluation, we sample 100K points $\mathbf{S} = \{\mathbf{s}_i\}$ uniformly distributed over the surface of a shape, and then use Chamfer Distance (CD) and F-score (F1) to evaluate the overall quality of a reconstructed mesh, and Normal Consistency (NC) to evaluate the quality of its surface normals.

Inspired by BSP-Net \cite{BSPNET}, we employ Edge Chamfer Distance (ECD) and Edge F-score (EF1) to evaluate the preservation of sharp edges. We use the same ``sharpness'' definition in BSP-Net as $\sigma(\mathbf{s}_i) = \min_{j \in \mathcal{N}_\varepsilon(\mathbf{s}_i)} |\mathbf{n}_i \cdot \mathbf{n}_j|$, where $\mathcal{N}_\varepsilon(\mathbf{s})$ extracts the indices of the points in $\mathbf{S}$ within distance $\varepsilon$ from $\mathbf{s}$, and $\mathbf{n}_i$ is the surface normal at point $\mathbf{s}_i$. We compute an ``edge sampling'' of the surface by retaining points for which $\sigma(\mathbf{s}_i){<}0.2$. Given two shapes, the ECD and EF1 between them are simply the CD and F1 between the corresponding edge samples.
We also count the number of generated vertices (\#V) and triangles (\#T) to reveal the fidelity-complexity trade-off.

\begin{figure*}
  \centering
  \includegraphics[width=0.95\linewidth]{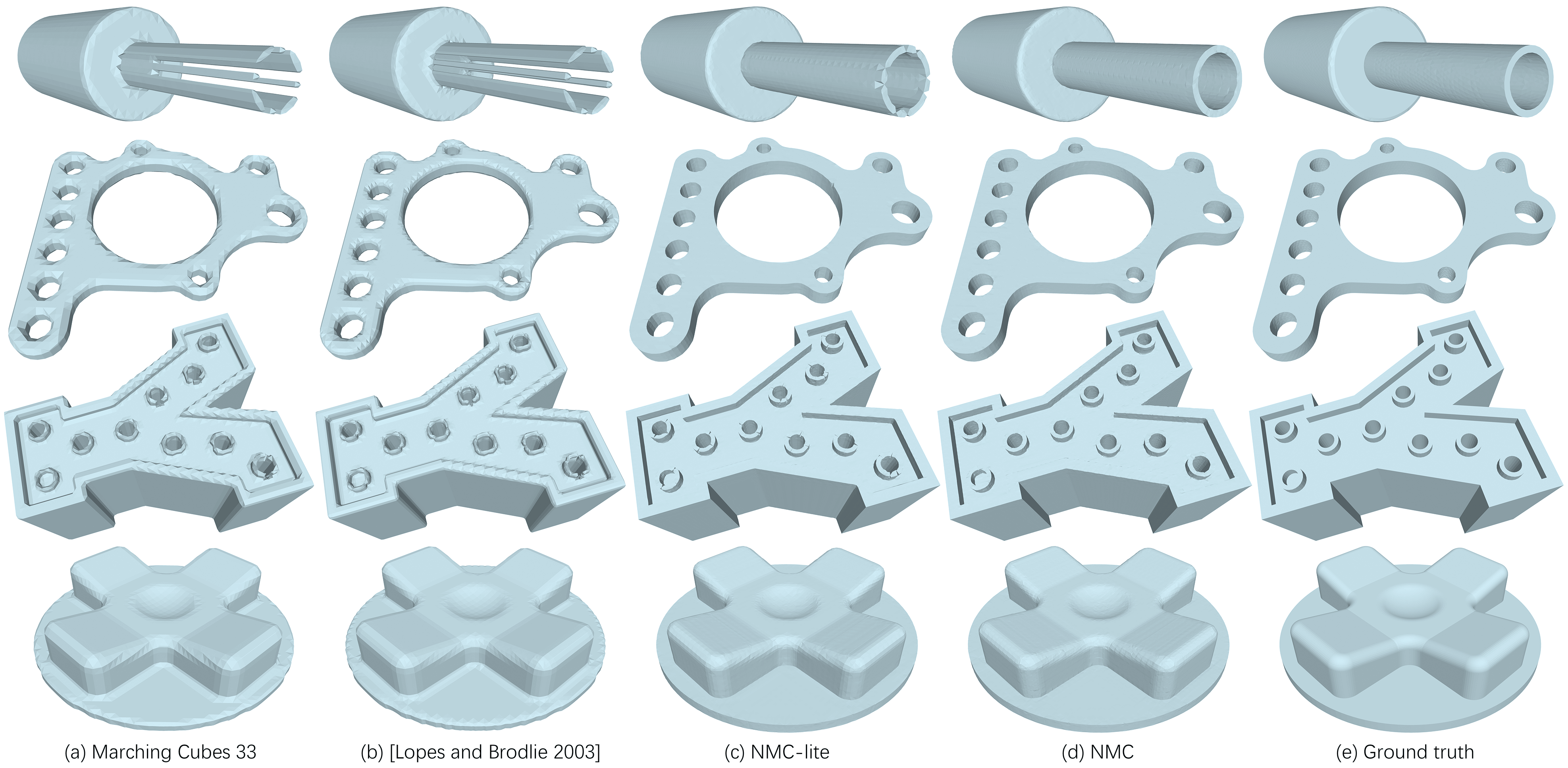}
	\caption{Results of reconstructing 3D meshes from SDF grid inputs at $64^3$ resolution. The shapes in the first two rows are from the ABC test set, and the last two rows from Thingi10K. More results and their mesh tessellations can be found in the supplementary material.}
	\label{fig:result_sdf}
\end{figure*}
\begin{table}
\caption{Quantitative comparison results on ABC test set with SDF input.}
\label{tab:compare_sdf_abc}
\begin{center}
\resizebox{1.0\linewidth}{!}{
\begin{tabular}{lrrrrrrr}
\toprule
$128^3$ resolution & CD{\small($\times 10^5$)}$\downarrow$ & F1$\uparrow$ & NC$\uparrow$ & ECD{\small($\times 10^2$)}$\downarrow$ & EF1$\uparrow$ & \#V & \#T \\
\midrule
MC33  & 4.143 & 0.870 & 0.972 & 4.063 & 0.193 & 22,048.41 & 44,107.11 \\
\toprule
$64^3$ resolution & CD{\small($\times 10^5$)}$\downarrow$ & F1$\uparrow$ & NC$\uparrow$ & ECD{\small($\times 10^2$)}$\downarrow$ & EF1$\uparrow$ & \#V & \#T \\
\midrule
MC33       & 4.850 & 0.788 & 0.950 & 5.736 & 0.105 &  5,472.51 & 10,953.67 \\
Lopes2003  & 4.803 & 0.798 & 0.958 & 6.841 & 0.100 & 21,979.95 & 43,892.05 \\
Trilinear  & 4.733 & 0.803 & 0.960 & 7.275 & 0.098 & - & - \\
NMC-lite   & 4.341 & {\bf 0.877} & {\bf 0.975} & {\bf 0.382} & {\bf 0.759} & 22,710.56 & 43,876.87 \\
NMC        & {\bf 4.323} & {\bf 0.877} & {\bf 0.975} & 0.390 & 0.758 & 42,766.54 & 85,543.83 \\
\toprule
$32^3$ resolution & CD{\small($\times 10^4$)}$\downarrow$ & F1$\uparrow$ & NC$\uparrow$ & ECD{\small($\times 10^2$)}$\downarrow$ & EF1$\uparrow$ & \#V & \#T \\
\midrule
MC33       & 5.239 & 0.570 & 0.900 & 5.504 & 0.048 & 1,297.38 & 2,595.47\\
Lopes2003  & 5.343 & 0.577 & 0.911 & 6.213 & 0.047 & 5,215.12 & 10,397.68 \\
Trilinear  & 5.161 & 0.585 & 0.915 & 7.217 & 0.045 & - & - \\
NMC-lite   & 3.922 & 0.823 & {\bf 0.950} & {\bf 0.532} & 0.631 & 5,464.48 & 10,389.43 \\
NMC        & {\bf 3.919} & {\bf 0.824} & 0.949 & 0.598 & {\bf 0.634} & 9,728.20 & 19,460.09 \\
\bottomrule
\end{tabular}
}
\end{center}
\bigskip\centering
\end{table}

\begin{figure}
  \centering
  \includegraphics[width=0.95\linewidth]{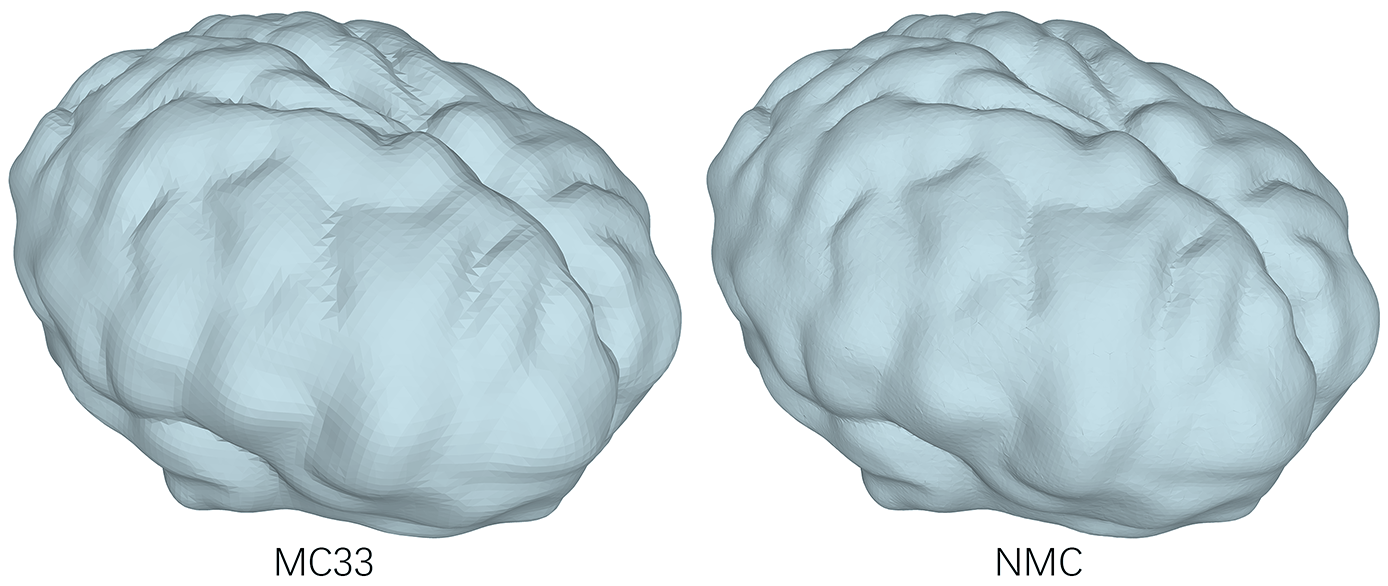}
	\caption{Reconstruction results on a brain model (in Thingi10K) with smooth features by MC33 and NMC, from SDF inputs. NMC preserves the surface details (especially around the valley areas) better.}
	\label{fig:smooth_brain}
\end{figure}
\begin{table}
\caption{Quantitative comparison on organic FAUST shapes with SDF input.}
\label{tab:compare_sdf_faust}
\begin{center}
\resizebox{1.0\linewidth}{!}{
\begin{tabular}{lrrrrrrr}
\toprule
$128^3$ resolution & CD{\small($\times 10^5$)}$\downarrow$ & F1$\uparrow$ & NC$\uparrow$ & ECD{\small($\times 10^2$)}$\downarrow$ & EF1$\uparrow$ & \#V & \#T \\
\midrule
MC33       & 4.533 & 0.985 & 0.984 & 0.892 & 0.383 & 12,551.21 & 25,076.50 \\
Lopes2003  & 4.487 & 0.985 & 0.986 & 0.858 & 0.409 & 50,649.41 & 100,417.26 \\
NMC-lite   & {\bf 3.696} & {\bf 0.992} & {\bf 0.987} & {\bf 0.559} & {\bf 0.628} & 50,205.72 & 100,401.08 \\
NMC        & 3.706 & {\bf 0.992} & {\bf 0.987} & 0.625 & {\bf 0.628} & 83,023.47 & 166,036.10 \\
\bottomrule
\end{tabular}
}
\end{center}
\bigskip\centering
\end{table}

\paragraph{Mesh extraction from SDF grids.}
We first compare the two versions of our method, NMC and NMC-lite, with the two best-known MC variants to date, Marching Cubes 33 \cite{lewiner2003efficient} (MC33) and \cite{lopes2003improving} (Lopes2003), on the task of mesh extraction from grids of SDF values.
Quantitative comparison results are reported in Table~\ref{tab:compare_sdf_abc}, with two choices of input resolutions: $64^3$ and $32^3$,
on the ABC test set.
The results show that, with the same SDF inputs, our method outperforms MC33 and Lopes2003 on all the quantitative measures considered.

\begin{figure}
  \centering
  \includegraphics[width=1.0\linewidth]{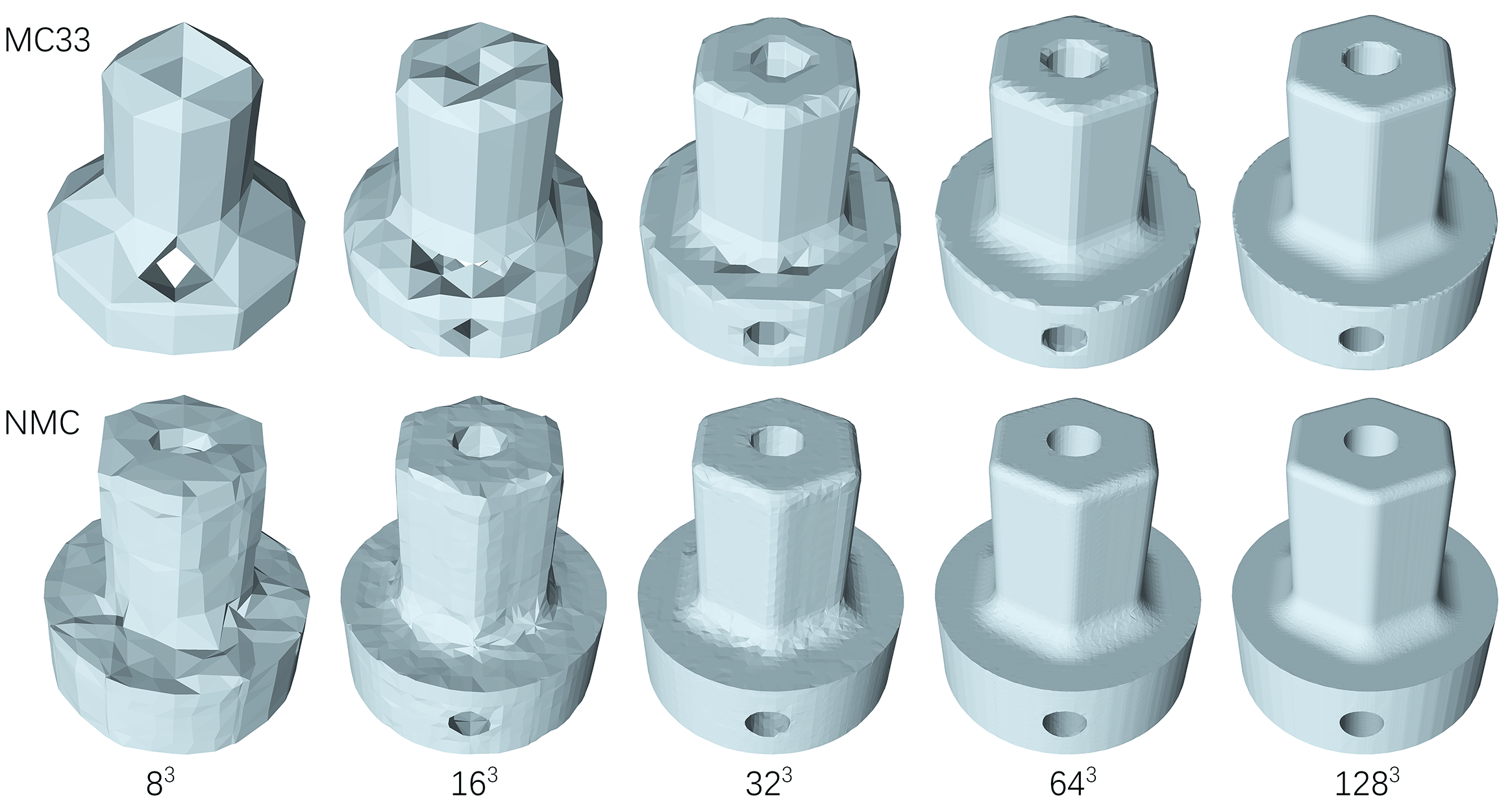}
	\caption{Results of reconstructing 3D mesh shapes from SDF inputs as the input grid resolutions vary. The holes in the MC33 results are due to incorrectly predicted topological cases. NMC consistently outforms MC33 at all input resolutions, up to $128^3$, but with a ``diminishing margin".}
	\label{fig:different_scale}
\end{figure}

We also add a row (top row in Table~\ref{tab:compare_sdf_abc}) for MC33 with the input resolution upscaled to $128^3$. As we can see, even with $8\times$ the amount of input information as NMC and NMC-lite, MC33 {\em underperforms\/} on all measures except for CD. In terms of edge preservation, our method is superior. This is also verified by the visual results in Figure~\ref{fig:different_scale}, comparing MC33 on $128^3$ input and NMC on $64^3$ input.

Figure~\ref{fig:result_sdf} shows qualitative comparisons between the various methods, on sample inputs from the ABC test set and Thingi10K. We can observe that NMC, and to a lesser extent, NMC-lite, are the only methods
that can recover sharp edges and corners, while the smooth curves are also well preserved. 
In fact, our method can reconstruct both sharp and soft edges well, as demonstrated in the last row.
Furthermore, examples in the first row and the third row (at a smaller scale) exhibit thin structures in a shape, which causes
both MC33 and Lobes2003 to produce incorrect local topologies, due to the trilinear interpolant assumption. On the other hand, our method
infers the correct topological cases --- the ambiguous Case 10.1.1 (see Figure~\ref{fig:3D_cube_tess_ours}), resulting in more faithful reconstructions.

In Figure~\ref{fig:teaser}(c), we show the isosurface of a trilinear interpolant, and in Table~\ref{tab:compare_sdf_abc}, we report quantitative results associated with trilinear interpolation. The ``ground truth'' trilinear interpolant could be considered as the upper bound of all methods that follow the trilinearity assumption. Therefore, our method outperforming the trilinear interpolant further proves that NMC is fundamentally superior at feature-preserving isosurface extraction.

\paragraph{Organic shapes.}
In Figure~\ref{fig:smooth_brain}, we show that when the ground truth shape has smooth undulations, our method is still able to reconstruct the surface details better than MC33.
For a more comprehensive test on organic shapes, we compare the various methods on 100 meshes of human body shapes from the FAUST dataset~\cite{FAUST}. The quantitative results in Table~\ref{tab:compare_sdf_faust} show that NMC and its variant can learn to predict both smooth and sharp features well, outperforming both MC33 and Lopes2003. Augmenting the training set with more organic shapes should further improve performance on such inputs, since our method is data-driven.

\paragraph{Varying input grid resolutions.}
In Figure~\ref{fig:different_scale}, we show how MC33 and NMC perform as the input SDF resolutions vary from $8^3$ to $128^3$, where we recall that our network was trained on meshes obtained at $32^3$ and $64^3$ resolutions. It is clear that our method can easily adapt to higher-resolution inputs, but degrades in reconstruction quality at the lower end. 
This is expected since as the cube size grows relative to the shape, the geometric varieties inside the cubes would surpass the set of topological cases covered by NMC. Nevertheless, NMC appears to consistently outperform MC33 at all resolutions, up to at least $128^3$. As the resolution continues to grow however, the difference between NMC and MC33 will diminish since the topological cases per cube would be much simplified.

\begin{figure*}
  \centering
  \includegraphics[width=0.95\linewidth]{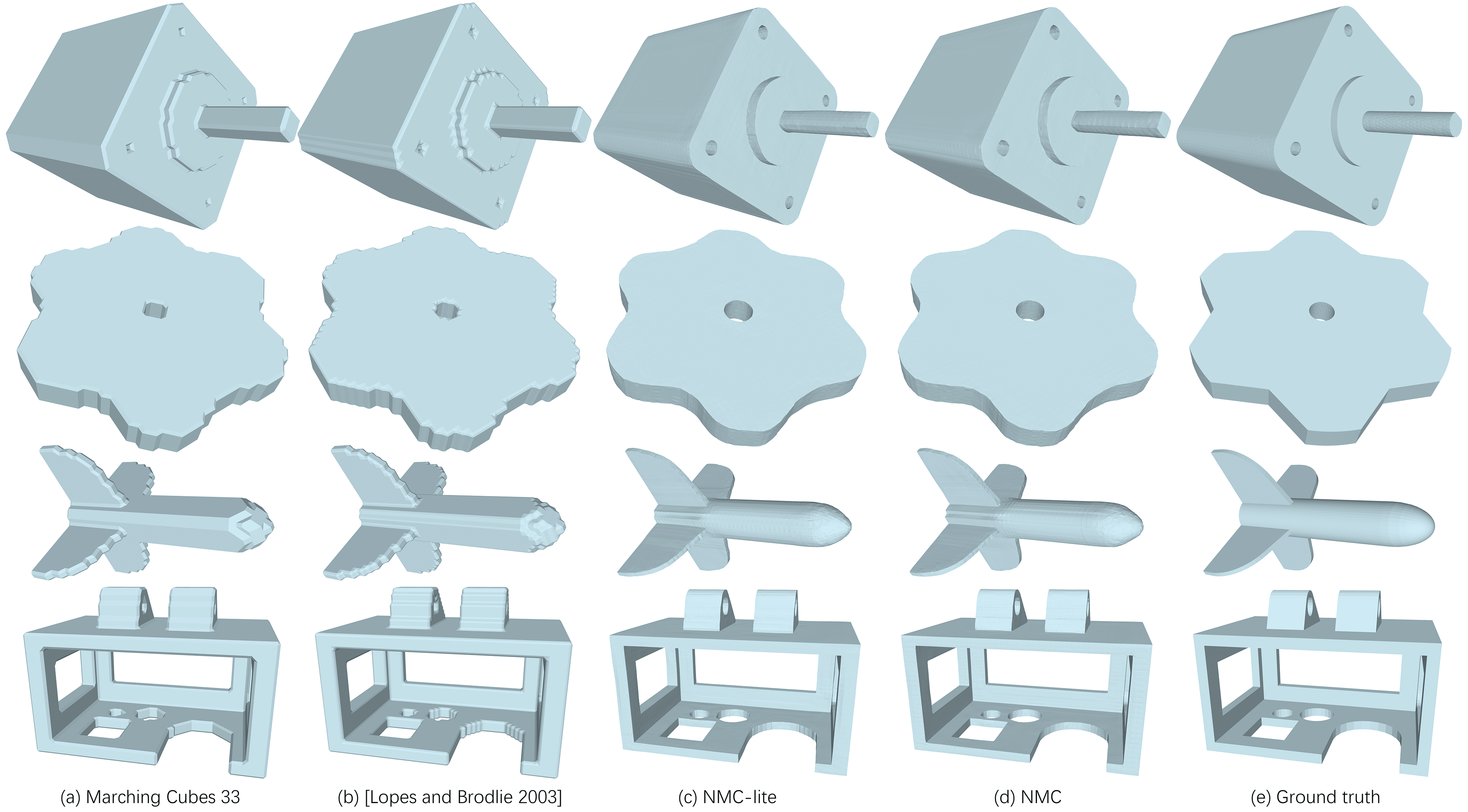}
	\caption{Results of reconstructing 3D meshes from binary voxel/occupancy inputs at $64^3$ resolution. The shapes in the first two rows are from the ABC test set, and the last two rows from Thingi10K. More results and their mesh tessellations can be found in the supplementary material.}
	\label{fig:result_voxel}
\end{figure*}
\begin{table}
\caption{Quantitative comparisons on ABC test set with binary voxel input.}
\label{tab:compare_voxel_abc}
\begin{center}
\resizebox{1.0\linewidth}{!}{
\begin{tabular}{lrrrrrrr}
\toprule
$64^3$ resolution & CD{\small($\times 10^5$)}$\downarrow$ & F1$\uparrow$ & NC$\uparrow$ & ECD{\small($\times 10^2$)}$\downarrow$ & EF1$\uparrow$ & \#V & \#T \\
\midrule
MC33       & 26.860 & 0.085 & 0.921 & 11.196 & 0.018 & 5,826.08 & 11,655.52 \\
Lopes2003  & 26.829 & 0.084 & 0.921 & 14.601 & 0.017 & 23,302.73 & 46,608.90 \\
Trilinear  & 26.826 & 0.084 & 0.921 & 14.866 & 0.017 & - & - \\
NMC-lite   & {\bf 9.302} & {\bf 0.443} & 0.930 & 0.559 & {\bf 0.365} & 22,185.94 & 42,915.64 \\
NMC        & 9.341 & 0.438 & {\bf 0.931} & {\bf 0.528} & 0.356 & 42,043.03 & 84,087.85 \\
\toprule
$32^3$ resolution & CD{\small($\times 10^4$)}$\downarrow$ & F1$\uparrow$ & NC$\uparrow$ & ECD{\small($\times 10^2$)}$\downarrow$ & EF1$\uparrow$ & \#V & \#T \\
\midrule
MC33       & 9.636 & 0.036 & 0.882 & 11.764 & 0.018 & 1,532.70 & 3,065.30 \\
Lopes2003  & 9.632 & 0.036 & 0.883 & 14.723 & 0.017 & 6,130.84 & 12,261.58 \\
Trilinear  & 9.641 & 0.035 & {\bf 0.884} & 14.820 & 0.017 & - & - \\
NMC-lite   & {\bf 5.909} & {\bf 0.237} & 0.871 & {\bf 0.901} & {\bf 0.112} & 5,236.79 & 9,975.67\\
NMC        & 6.029 & 0.232 & 0.871 & 0.910 & 0.109 & 9,469.84 & 18,933.65 \\
\bottomrule
\end{tabular}
}
\end{center}
\bigskip\centering
\vspace{-10pt}
\end{table}

\paragraph{Mesh extraction from binary voxels.}
When the inputs are binary voxels instead of signed distances, the isosurface extraction task becomes significantly more difficult, since voxel occupancies possess considerably less information. Not only are the point-to-surface distances lost in the occupancies, but the signs could also be inaccurate: a point outside the shape in the SDF grid may become ``inside'' in the voxel grid. One can observe a quality drop from the visual results shown in Figure~\ref{fig:result_voxel}. Even our method cannot always produce faithful reconstructions due to the ambiguities, e.g., the rod in the first row could be rectangular or circular, and the edges in the second row could be smooth or sharp - both would yield identical voxel grids. However, our learning-based approach is able to narrow down the possible geometries and topologies by observing local neighborhoods. 
As shown by the quantitative results in Table~\ref{tab:compare_voxel_abc}, our method outperforms other MC variants and the trilinear interpolant on all measures, except for NC, by a substantial margin.

\paragraph{Generalizability.}
To demonstrate generalizability of our networks, which were trained on ABC, we test them on the first 2,000 valid shapes from Thingi10K, using exactly the same network weights as those in the previous experiments. Tables~\ref{tab:compare_sdf_thingi10k} and~\ref{tab:compare_voxel_thingi10k} show quantitative comparison results on SDF and binary voxel inputs, respectively. Some qualitative results are given in Figures~\ref{fig:result_sdf} and~\ref{fig:result_voxel}. We can observe a similar performance boost over the other methods in comparison. Note however that in Table~\ref{tab:compare_voxel_thingi10k}, our method does not significantly outperform other methods at the $32^3$ input voxel resolution. This may be due to NMC being overfit to the ABC training set, since the shape resolution ($32^3$) is getting close to the size of the receptive field of our voxel processing network ($15^3$).

\begin{figure}
  \centering
  \includegraphics[width=1.0\linewidth]{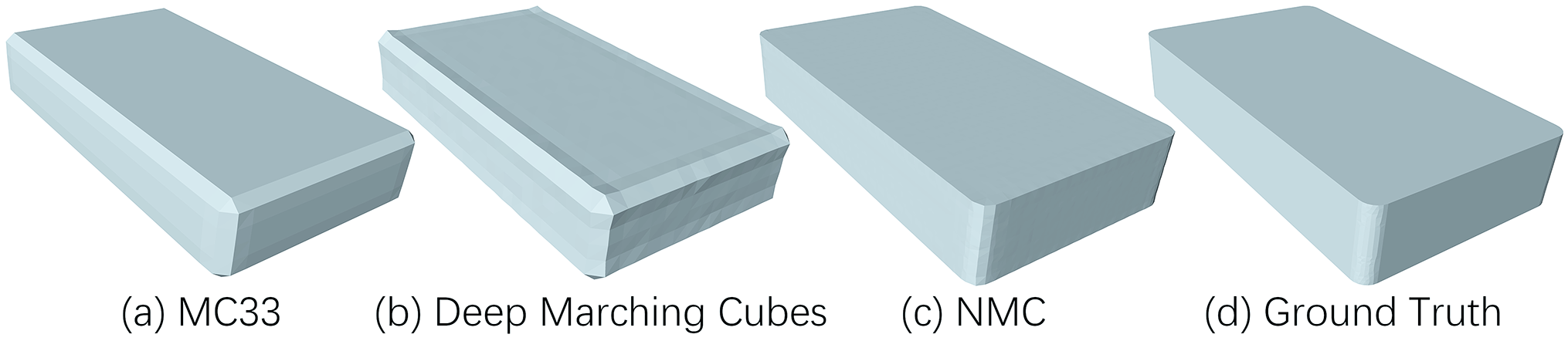}
	\caption{Comparing NMC with MC33 and Deep Marching Cubes (DMC) \cite{DeepMarchingCubes} on feature preservation.}
	\label{fig:compare_dmc}
\end{figure}

\paragraph{Comparison to DMC}
In Figure~\ref{fig:compare_dmc}, we compare NMC with DMC \cite{DeepMarchingCubes} on feature-preserving mesh reconstruction.
Since the network architecture of DMC is not designed to perform general isosurface extraction, we train their network to {\em overfit\/} on a single input shape with 65,536 uniformly sampled points as supervision. As we can see, even with such an overfitting, DMC is still unable to recover sharp features, which is mainly due to its adoption of only few classical MC tessellations representing simple topologies. Related to this, while DMC is trained to minimize point-to-triangle distances, it does not provide the tessellations to support sharp edges. As a result, the reconstructed geometry near edges is ``bulging'' in order to minimize distances 
to the training points.

\begin{table}
\caption{Quantitative comparison results on Thingi10K with SDF input.}
\label{tab:compare_sdf_thingi10k}
\begin{center}
\resizebox{1.0\linewidth}{!}{
\begin{tabular}{lrrrrrrr}
\toprule
$64^3$ resolution & CD{\small($\times 10^5$)}$\downarrow$ & F1$\uparrow$ & NC$\uparrow$ & ECD{\small($\times 10^2$)}$\downarrow$ & EF1$\uparrow$ & \#V & \#T \\
\midrule
MC33       & 3.195 & 0.795 & 0.945 & 3.763 & 0.099 & 5,517.51 & 11,044.35 \\
Lopes2003  & 3.084 & 0.805 & 0.953 & 4.567 & 0.087 & 22,224.23 & 44,135.98 \\
Trilinear  & 3.076 & 0.811 & 0.956 & 5.211 & 0.084 & - & - \\
NMC-lite   & {\bf 2.470} & {\bf 0.893} & {\bf 0.972} & 0.330 & {\bf 0.722} & 22,991.80 & 44,109.17\\
NMC        & 2.477 & {\bf 0.893} & {\bf 0.972} & {\bf 0.312} & {\bf 0.722} & 40,951.73 & 81,910.41\\
\toprule
$32^3$ resolution & CD{\small($\times 10^4$)}$\downarrow$ & F1$\uparrow$ & NC$\uparrow$ & ECD{\small($\times 10^2$)}$\downarrow$ & EF1$\uparrow$ & \#V & \#T \\
\midrule
MC33       & 10.519 & 0.540 & 0.882 & 4.046 & 0.040 & 1,284.98 & 2,569.73 \\
Lopes2003  & 10.473 & 0.547 & 0.893 & 4.596 & 0.038 & 5,163.28 & 10,281.15 \\
Trilinear  & 10.431 & 0.555 & 0.897 & 5.180 & 0.037 & - & - \\
NMC-lite   & {\bf 8.425} & 0.807 & {\bf 0.935} & 0.600 & {\bf 0.542} & 5,423.92 & 10,263.13 \\
NMC        & 8.454 & {\bf 0.808} & 0.933 & {\bf 0.596} & 0.539 & 9,161.94 & 18,327.88 \\
\bottomrule
\end{tabular}
}
\end{center}
\bigskip\centering
\end{table}

\begin{table}
\caption{Quantitative comparisons on Thingi10K with binary voxel input.}
\label{tab:compare_voxel_thingi10k}
\begin{center}
\resizebox{1.0\linewidth}{!}{
\begin{tabular}{lrrrrrrr}
\toprule
$64^3$ resolution & CD{\small($\times 10^5$)}$\downarrow$ & F1$\uparrow$ & NC$\uparrow$ & ECD{\small($\times 10^2$)}$\downarrow$ & EF1$\uparrow$ & \#V & \#T \\
\midrule
MC33       & 25.538 & 0.069 & 0.907 & 7.411 & 0.017 & 5,939.62 & 11,881.67 \\
Lopes2003  & 25.526 & 0.068 & 0.908 & 11.948 & 0.015 & 23,757.44 & 47,517.48 \\
Trilinear  & 25.510 & 0.068 & 0.909 & 12.598 & 0.015 & - & - \\
NMC-lite   & {\bf 6.055} & {\bf 0.495} & {\bf 0.923} & 0.606 & {\bf 0.328} & 22,540.88 & 43,272.05 \\
NMC        & 6.108 & 0.493 & {\bf 0.923} & {\bf 0.602} & 0.314 & 40,430.06 & 80,861.75 \\
\toprule
$32^3$ resolution & CD{\small($\times 10^4$)}$\downarrow$ & F1$\uparrow$ & NC$\uparrow$ & ECD{\small($\times 10^2$)}$\downarrow$ & EF1$\uparrow$ & \#V & \#T \\
\midrule
MC33       & 9.247 & 0.028 & 0.865 & 8.632 & 0.017 & 1,553.93 & 3,107.50 \\
Lopes2003  & {\bf 9.246} & 0.028 & {\bf 0.867} & 12.344 & 0.015 & 6,215.99 & 12,431.69 \\
Trilinear  & 9.256 & 0.028 & {\bf 0.867} & 12.709 & 0.015 & - & - \\
NMC-lite   & 9.998 & {\bf 0.258} & 0.852 & {\bf 0.946} & {\bf 0.096} & 5,261.82 & 9,971.62 \\
NMC        & 10.177 & 0.256 & 0.852 & 0.957 & 0.093 & 9,043.78 & 18,083.90\\
\bottomrule
\end{tabular}
}
\end{center}
\bigskip\centering
\end{table}

\paragraph{Input and output complexities.}
When making comparisons, the input resolutions to all methods are identical, but the output complexities do vary, as shown in Tables~\ref{tab:compare_sdf_abc}-\ref{tab:compare_voxel_thingi10k}, in terms of the average triangle and vertex counts. With the
same tessellation templates, hence comparable output complexities, NMC-lite outperforms Lopes2003 on all fronts, both 
quantitatively and qualitatively, offering clear evidence for the effectiveness of our data-driven approach for isosurfacing. The new
tessellation templates designed for NMC are more refined, resulting in higher triangle counts, but also improved reconstruction quality, as
shown in Figures~\ref{fig:result_sdf} and~\ref{fig:result_voxel}.

\paragraph{NMC vs~NMC-lite.}
Quantitatively, the performances of NMC and NMC-lite are quite similar, proving the robustness of our representation design. However, the visual quality of NMC results tends to be better than that of NMC-lite, at the expense of almost doubling the triangle counts. Thus, if a lower output complexity is desired, one may choose NMC-lite over NMC. But since NMC-lite employs simpler tessellation templates, it may not be able to recover specific fine shape features, such as the thin structures in the examples from the first and third rows of Figure~\ref{fig:result_sdf}, where the cubes with Case 3.2 were not well reconstructed. Also, we have observed that the {\em triangle quality\/} resulting from NMC is generally better than that from NMC-lite (e.g., see Figure~\ref{fig:teaser}), since
the NMC tessellation templates were designed to better avoid thin/sliver triangles.

\paragraph{Training and testing times.}
Network training takes about 3 days on one Nvidia RTX 2080 Ti GPU for SDF processing and 2 days for binary voxel inputs.
We tested inference time on the entire ABC test set with $64^3$ inputs: the average is $0.006$ second per shape for MC33 (implemented in scikit-image~\cite{van2014scikit}), and $0.762$ second for NMC (with $0.719$s spent on network forwarding in PyTorch~\cite{pytorch} and $0.042$s for meshing in Cython~\cite{cython}). Note that currently, our network is still quite crude, as we prioritize accuracy over speed. Possible speed-up could be achieved via neural architecture search and network pruning.

\begin{figure*}
  \centering
  \includegraphics[width=0.95\linewidth]{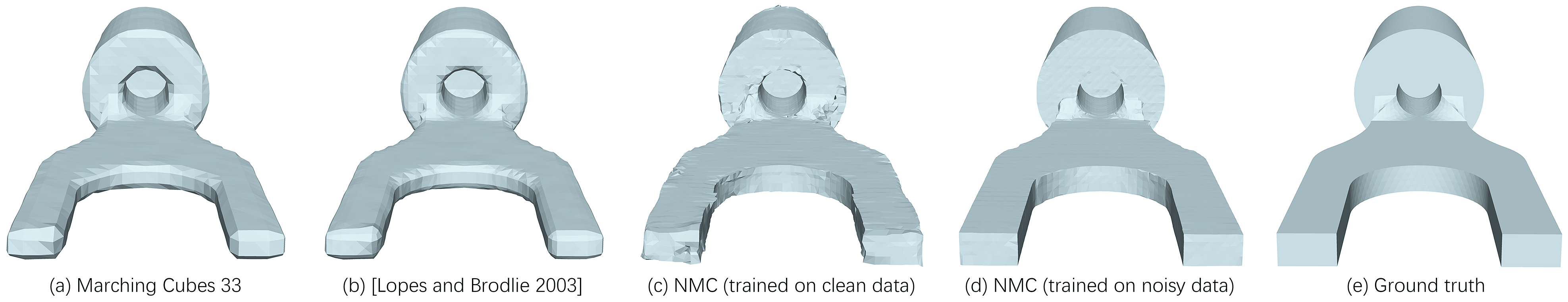}
	\caption{Results of reconstructing 3D meshes from a {\em noisy\/} SDF grid input at $64^3$ resolution.}
	\label{fig:result_noisy_sdf}
\end{figure*}

\paragraph{Noisy inputs.}
Finally, we show that NMC can also learn to extract {\em clean\/} meshes from noisy grid inputs when the network is trained on such data, such as those generated by current neural implicit models~\cite{IMNET,DeepSDF,OccNet}. To test this capability, we run a state-of-the-art neural implicit model, SIREN \cite{sitzmann2019siren}, on the ABC dataset to fit each shape, but with only 4,096 training points per shape. The sparsity of the training points makes the output implicit fields necessarily noisy, as can be observed from Figures~\ref{fig:result_noisy_sdf}(a-b). 

In our previous experiments, we trained NMC on clean data from ABC and assumed that the testing inputs were also clean. A model trained this way may fail when the input is noisy, as shown in Figure~\ref{fig:result_noisy_sdf}(c). To remedy this, we divide the noisy inputs into 80\% training and 20\% testing as before, and use the noisy inputs to train the NMC model from scratch. The re-trained NMC improves significantly on inputs from the noisy test set, as shown in Figure~\ref{fig:result_noisy_sdf}, demonstrating that our method can adapt to different inputs (such as voxels and noisy data), if trained on them. 

Table~\ref{tab:compare_noisy_sdf_abc} shows a quantitative comparison involving NMC trained on clean vs.~noisy data.
We note that Chamfer Distance (CD) is rather sensitive to outliers, e.g., SIREN may generate blobs in the empty region that are far away from the shape, which can impact CD heavily. In comparison, F1 is a more robust quality measure to outliers, as discussed in \cite{tatarchenko2019single}.

\begin{table}
\caption{Quantitative comparison on ABC test set with noisy SDF input.}
\label{tab:compare_noisy_sdf_abc}
\begin{center}
\resizebox{1.0\linewidth}{!}{
\begin{tabular}{lrrrrr}
\toprule
$64^3$ resolution & CD{\small($\times 10^5$)}$\downarrow$ & F1$\uparrow$ & NC$\uparrow$ & ECD{\small($\times 10^2$)}$\downarrow$ & EF1$\uparrow$ \\
\midrule
MC33                        & 16.611 & 0.710 & 0.942 & 3.360 & 0.100 \\
Lopes2003                   & 16.545 & 0.714 & 0.947 & 3.692 & 0.093 \\
NMC (trained on clean data) & {\bf 15.340} & 0.769 & 0.941 & 0.574 & 0.502 \\
NMC (trained on noisy data) & 15.627 & {\bf 0.802} & {\bf 0.951} & {\bf 0.359} & {\bf 0.640} \\
\bottomrule
\end{tabular}
}
\end{center}
\bigskip\centering
\end{table}

\section{Conclusion, limitation, and future work}
\label{sec:future}

\begin{figure}
  \centering
  \includegraphics[width=1.0\linewidth]{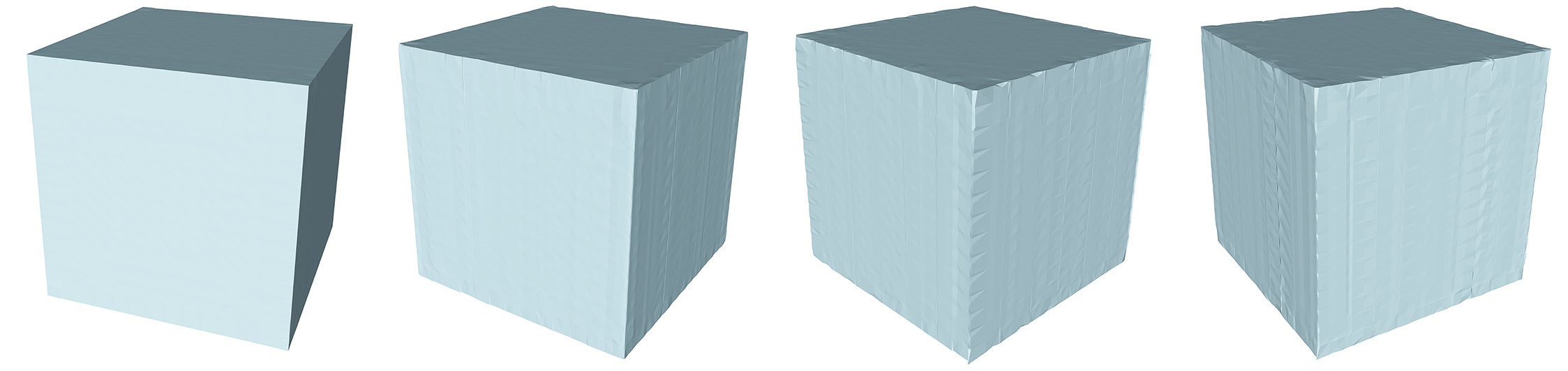}
	\caption{NMC may produce artifacts when the input is oriented at an ``unusual'' angle relative to the training shapes. From left to right: reconstructions of a cube that is axis-aligned, then rotated by $\frac{1}{14}\pi$, $\frac{2}{14}\pi$, and $\frac{3}{14}\pi$.}
	\label{fig:limitation_rotation}
\end{figure}
\begin{figure}
  \centering
  \includegraphics[width=0.9\linewidth]{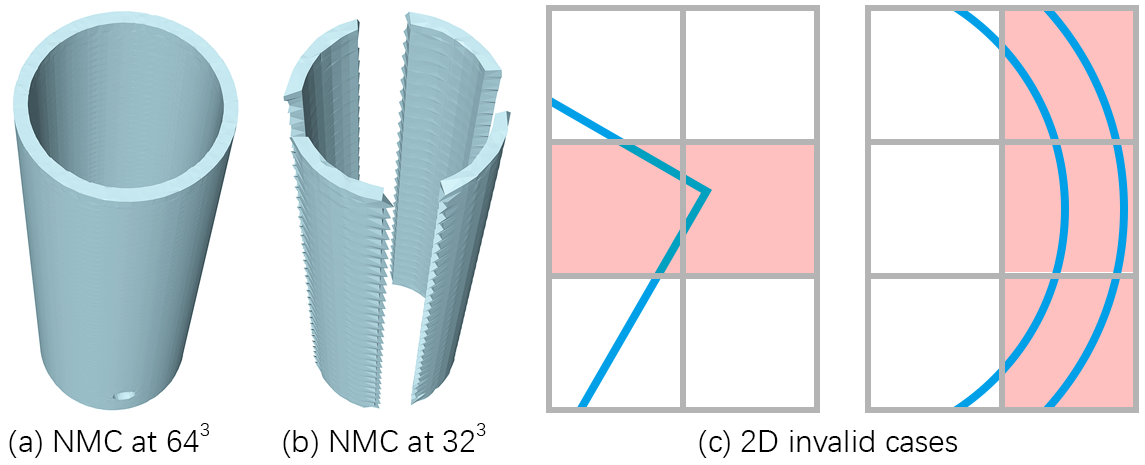}
	\caption{NMC cannot account for certain topological cases (deemed ``invalid''), e.g., multiple intersections along an edge as highlighted in red (c). The reconstruction failure in (b) is due to similar invalid cases in 3D.}
	\label{fig:limitation_cases}
\end{figure}

In this paper, we show for the first time that the mesh reconstruction quality by Marching Cubes (MC), one of the most classical algorithms in computer graphics, can be significantly boosted by machine learning. In Neural Marching Cubes (NMC), we introduce the first MC-based 
approach capable of recovering sharp geometric features without requiring additional inputs, such as normal information.
Trained on automatically generated ``ground-truth'' meshes,
our method shows superior performance in preserving various geometric features such as sharp/soft edges, corners, thin structures, etc., compared to other isosurfacing algorithms that take uniform grids of signed distances or
binary occupancies as inputs.
We also designed an efficient parameterization to represent a triangle mesh, compatible with neural processing, so that our NMC network can directly output the meshes without postprocessing. 

The main limitation of our method in terms of isosurfacing is its sensitivity to rotation, as shown in Figure~\ref{fig:limitation_rotation}. This is mainly due to the dataset we train the network on, as the shapes in the ABC dataset are mostly aligned with the coordinate axes.
This data characteristic also motivated our definition of the smoothness term in Eq.~(\ref{eq:smoothness}).
Performing random rotation augmentation on the training data is a viable solution, but would require longer training time and larger networks to fit. Second, as we follow the common assumption in MC that if the two vertices of a cube edge have different signs, then there is one and only one intersection point, several topological cases (as shown in Figure~\ref{fig:limitation_cases}) cannot be represented. Adding more intersection points should cover most of such cases, and the numbers and the positions of the edge vertices can be learned from data.

Another limitation is that we do not have a built-in mechanism to avoid self-intersections 
in the output meshes.
When testing on $64^3$ SDF inputs from ABC, $32.7$\% of the meshes produced by NMC contain self-intersections, but they involve only $0.0086$\% of the triangles, which translate to about $7.39$ triangles or around two separate intersections, per shape. For NMC-lite, the corresponding numbers are $29.6$\%, $0.0078$\%, and $3.40$, respectively.
Most of the intersections happen in cases where the structure in a cube cannot be represented by our tessellation templates, such as those in Figure~\ref{fig:limitation_cases}(b).
Last but not the least, our current model does not allow the learning of an arbitrary {\em tessellation style\/}, e.g., meshes whose triangles are all close to being equilaterals. The challenge is on how to prepare the proper training meshes to work under our designed templates.

Our proposed representation is not constrained to isosurfacing. It is a general shape representation that can be adopted to other tasks such as shape upsampling and generation, pointing to a worthy direction for future work. On the other hand, even when the input is a uniform grid, the output mesh does not have to be uniform. A simple plane only requires a few triangles to model, but a curved surface needs more. Therefore, learning to adaptively allocate vertices and triangles according to feature complexity could yield more efficient algorithms and control the explosion of triangle counts in MC, as reported in BSP-NET \cite{BSPNET}.

\begin{acks}
We thank the reviewers for their valuable comments, and Thomas Funkhouser for helpful discussions. This work was supported in part by NSERC (no.~611370) and gift funds from Google and Adobe.
\end{acks}

\bibliographystyle{ACM-Reference-Format}
\bibliography{mainbib}

\appendix

\section{Comparison of different smoothness terms}
\label{sec:appendix}

\begin{figure}
  \centering
  \includegraphics[width=1.0\linewidth]{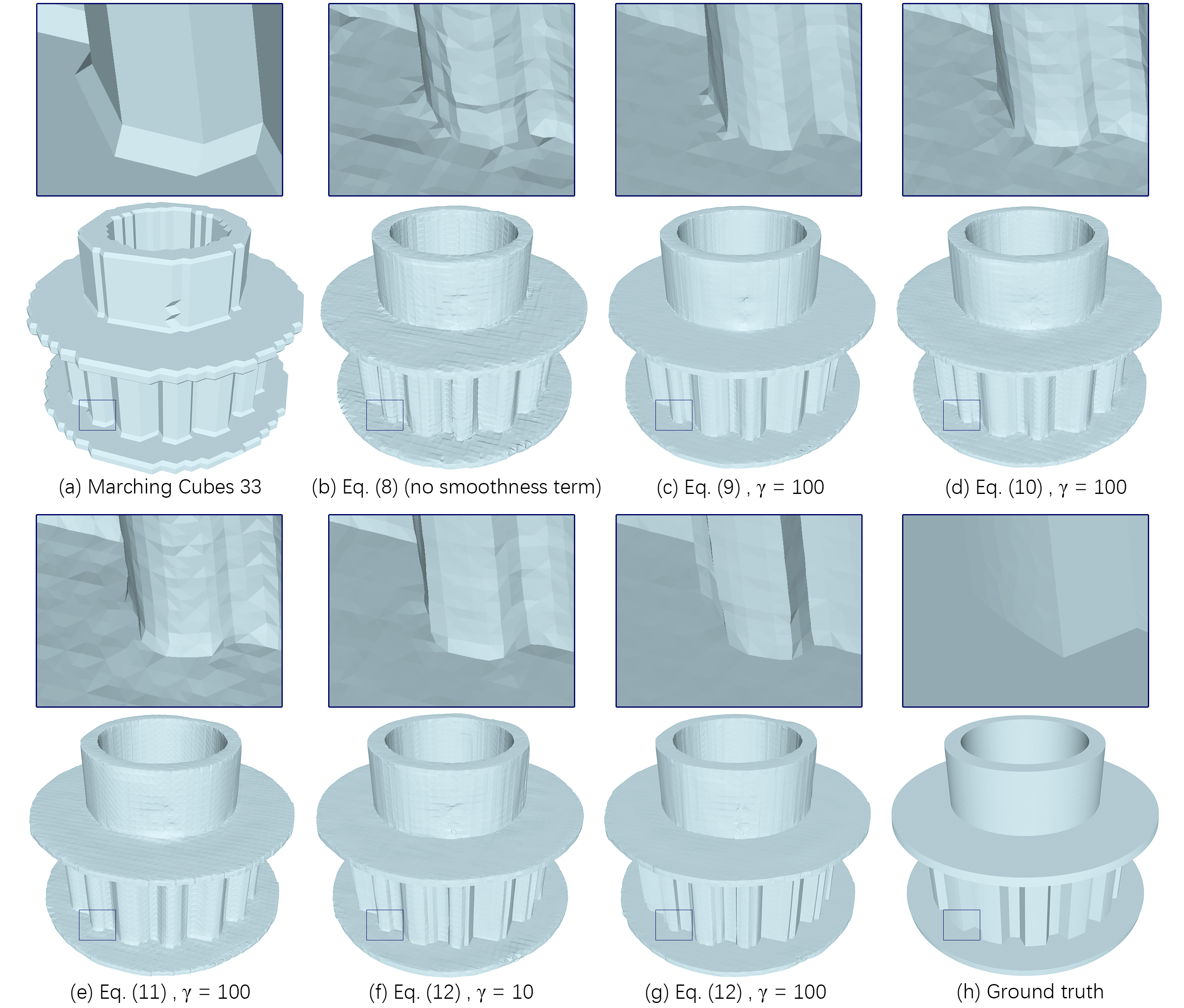}
	\caption{Visual comparisons of different smoothness terms on ABC test set with binary voxel input at $64^3$ resolution.}
	\label{fig:voxel_ABC_smoothness_small}
\end{figure}

\begin{table}
\caption{Comparison of different smoothness terms on the ABC test set. The underlined superscripts show rankings of the quantitative performances, where the overall best performing row is highlighted in bold.}
\label{tab:voxel_ABC_smoothness_small}
\begin{center}
\resizebox{1.0\linewidth}{!}{
\begin{tabular}{lrrrrr}
\toprule
$64^3$ resolution & CD{\small($\times 10^5$)}$\downarrow$ & F1$\uparrow$ & NC$\uparrow$ & ECD{\small($\times 10^2$)}$\downarrow$ & EF1$\uparrow$ \\
\midrule
Eq. (8)                               & $^{\underline 4}$9.355 & $^{\underline 1}$0.445 & $^{\underline 3}$0.932 & $^{\underline 5}$0.633 & $^{\underline 4}$0.328 \\
Eq. (9), $\gamma=100$    & $^{\underline 2}$9.329 & $^{\underline 3}$0.435 & $^{\underline 3}$0.932 & $^{\underline 4}$0.615 & $^{\underline 3}$0.353 \\
Eq. (10), $\gamma=100$  & $^{\underline 6}$9.539 & $^{\underline 3}$0.435 & $^{\underline 2}$0.933 & $^{\underline 3}$0.612 & $^{\underline 5}$0.320 \\
Eq. (11), $\gamma=100$   & $^{\underline 5}$9.518 & $^{\underline 5}$0.434 & $^{\underline 1}$0.934 & $^{\underline 2}$0.562 & $^{\underline 6}$0.313 \\
{\bf Eq. (12),} $\gamma${\bf =10}     & $^{\underline 3}$9.341 & $^{\underline 2}$0.438 & $^{\underline 5}$0.931 & $^{\underline 1}$0.528 & $^{\underline 2}$0.356 \\
Eq. (12), $\gamma=100$   & $^{\underline 1}$9.327 & $^{\underline 6}$0.427 & $^{\underline 6}$0.923 & $^{\underline 6}$0.669 & $^{\underline 1}$0.359 \\
\bottomrule
\end{tabular}
}
\end{center}
\bigskip\centering
\vspace{-10pt}
\end{table}

Due to considerable ambiguities in possible shapes represented by binary voxels, we need an extra smoothness term to regularize the generated surfaces. 
We reuse notations from Section~\ref{sec:method_nn} for $E_s$, $F^u_s$, and $H^u_s$, and $\mathbbm{1}(\cdot)$.
Further, let $F_s^{uv} = F_s^v - F_s^u$, $H_s^{uv} = H_s^v - H_s^u$,
$[F_s^{u}]_x$ be the $x$ coordinate of $F_s^{u}$, and
$[F_s^{uv}]_{yz} = \sqrt{[F_s^{uv}]_y^2+ [F_s^{uv}]_z^2}$. 
We have experimented with the following settings: 

\begin{equation}
\label{eq:smoothness1}
L_{smooth}^{(1)} = 0 \textrm{\;\;\;\;(No smoothness term)}.
\end{equation}

\begin{equation}
\label{eq:smoothness2}
L_{smooth}^{(2)} = \mathbb{E}_{s \sim \mathcal{S}} \sum_{(u,v) \in E_s} \big\| \;\; F_s^{uv} - H_s^{uv} \;\;\big\|_2^2.
\end{equation}

\begin{align}
\begin{aligned}
\label{eq:smoothness3}
L_{smooth}^{(3)} = \mathbb{E}_{s \sim \mathcal{S}} \sum_{(u,v) \in E_s} \big( L^{x}_{y} + L^{x}_{z} + L^{y}_{x} + L^{y}_{z} + L^{z}_{x} + L^{z}_{y} \big), \\
\textrm{where} \; L^{x}_{y} = \big( \;\;  [H_s^{uv}]_x \cdot \big| [F_s^{uv}]_y \big| - [F_s^{uv}]_x \cdot \big| [H_s^{uv}]_y \big| \;\;\big)^2.
\end{aligned}
\end{align}

\begin{align}
\begin{aligned}
\label{eq:smoothness4}
L_{smooth}^{(4)} = \mathbb{E}_{s \sim \mathcal{S}} \sum_{(u,v) \in E_s} \big( L^{x}_{yz} + L^{y}_{xz} + L^{z}_{xy} \big), \\
\textrm{where} \; L^{x}_{yz} = \big( \;\;  [H_s^{uv}]_x \cdot [F_s^{uv}]_{yz} - [F_s^{uv}]_x \cdot [H_s^{uv}]_{yz} \;\;\big)^2.
\end{aligned}
\end{align}

\begin{equation}
\label{eq:smoothness5}
L_{smooth}^{(5)} = \mathbb{E}_{s \sim \mathcal{S}} \sum_{(u,v) \in E_s} \big\| \;\; \mathbbm{1}(|F_s^{uv}|<\sigma) \cdot H_s^{uv} \;\;\big\|_2^2.
\end{equation}

The smoothness term $L_{smooth}^{(2)}$ in Eq.~(\ref{eq:smoothness2}) minimizes the differences between the edge gradients on the predicted mesh and those on the GT.
$L_{smooth}^{(3)}$ and $L_{smooth}^{(4)}$ try to preserve the absolute angles of the edges. In an ideal situation, $[H^{uv}]_x / |[H^{uv}]_y| = [F^{uv}]_x / |[F^{uv}]_y|$, therefore $[H^{uv}]_x \cdot |[F^{uv}]_y| - [F^{uv}]_x \cdot |[H^{uv}]_y| = 0$. Eq.~(\ref{eq:smoothness3}) minimizes the squared error of such terms. Eq.~(\ref{eq:smoothness4}) is a variant of Eq.~(\ref{eq:smoothness3}), while Eq.~(\ref{eq:smoothness5}) is equivalent to Eq.~(\ref{eq:smoothness}) in Section~\ref{sec:method_nn}. 

The performances of the different smoothness terms are shown in Table~\ref{tab:voxel_ABC_smoothness_small} and exhibited visually in Figure~\ref{fig:voxel_ABC_smoothness_small}. 
Based on the visual results and
overall quantitative performances, we have adopted $L_{smooth}^{(5)}$ in our work, with $\gamma=10$ and $\sigma = 0.0002$.

\end{document}